%% file: DAFx25_tmpl.tex
\def\papertitle{Physics-Informed Deep Learning for Nonlinear Friction Model of Bow-string Interaction}
\def\paperauthorA{Xinmeng Luan}
\def\paperauthorB{Gary Scavone}
\documentclass[twoside,a4paper]{article}
\usepackage{etoolbox}
\usepackage{dafx_25}
\usepackage{amsmath,amssymb,amsfonts,amsthm}
\usepackage{euscript}
\usepackage[T1]{fontenc}
\usepackage[utf8]{inputenc}
\usepackage{ifpdf}
\usepackage[english]{babel}
\usepackage{caption}
\usepackage{subfig} % or can use subcaption package
\usepackage{color}
\usepackage{tikz}
\usepackage{pgfplots}
\usepackage{amssymb}
\usepackage{siunitx}
\usepackage{subfig} 
\usepackage{setspace}
\input glyphtounicode
\ninept

\newcounter{numauth}
\newcounter{listcnt}
\newcommand\authcnt[1]{\ifdefined#1 \stepcounter{numauth} \fi}

\newcommand\addauth[1]{
\ifdefined#1 
\stepcounter{listcnt}
\ifnum \value{listcnt}<\value{numauth}
\appto\authorslist{, #1}
\else
\appto\authorslist{~and~#1}
\fi
\fi}
\authcnt{\paperauthorB}
\authcnt{\paperauthorC}
\authcnt{\paperauthorD}
\authcnt{\paperauthorE}
\authcnt{\paperauthorF}
\authcnt{\paperauthorG}
\authcnt{\paperauthorH}
\authcnt{\paperauthorI}
\authcnt{\paperauthorJ}
\def\authorslist{\paperauthorA}
\addauth{\paperauthorB}
\addauth{\paperauthorC}
\addauth{\paperauthorD}
\addauth{\paperauthorE}
\addauth{\paperauthorF}
\addauth{\paperauthorG}
\addauth{\paperauthorH}
\addauth{\paperauthorI}
\addauth{\paperauthorJ}
\usepackage{times}
\newif\ifpdf
\ifx\pdfoutput\relax
\else
   \ifcase\pdfoutput
      \pdffalse
   \else
      \pdftrue
   \fi
\fi

\usepackage{ifpdf}
\ifpdf
  \usepackage[
    pdftex,
    pdftitle={\papertitle},
    pdfauthor={\authorslist},
    pdfsubject={Proceedings of the 28th International Conference on Digital Audio Effects (DAFx25)},
    colorlinks=false,
    bookmarksnumbered,
    pdfstartview=XYZ
  ]{hyperref}
  \usepackage{graphicx} % load only once, no clash
\else
  \usepackage{graphicx} % load only once here
  \usepackage[dvips,
    pdftitle={\papertitle},
    pdfauthor={\authorslist},
    pdfsubject={Proceedings of the 28th International Conference on Digital Audio Effects (DAFx25)},
    colorlinks=false,
    bookmarksnumbered,
    pdfstartview=XYZ
  ]{hyperref}
\fi

\usepackage[hypcap=true]{caption}
\title{\papertitle}

\affiliation{
\paperauthorA\,and \paperauthorB \,\thanks{\vspace{-3mm}}}
{CAML\sthanks{Computational Acoustic Modeling Laboratory,  McGill University }, CIRMMT \sthanks{Center for Interdisciplinary Research in Music Media and Technology}, \\
{\tt xinmeng.luan@mail.mcgill.ca, gary.scavone@mcgill.ca
}
}

\begin{document}
% more pdf-tex settings:
\ifpdf % used graphic file format for pdflatex
  \DeclareGraphicsExtensions{.png,.jpg,.pdf}
\else  % used graphic file format for latex
  \DeclareGraphicsExtensions{.eps}
\fi

%\makeatletter
%\pdfbookmark[0]{\@pdftitle}{title}
%\makeatother

\maketitle

\begin{abstract}
% Physics-informed deep learning has recently gained recognition as a transformative approach for solving ordinary differential equations (ODEs) and partial differential equations (PDEs), achieving remarkable success in various engineering fields.  Two leading frameworks within this paradigm are Physics-Informed Neural Networks (PINNs) and Physics-Informed Deep Operator Networks (PI-DeepONets). In this study, we expand the application of PINNs and PI-DeepONets to the field of musical acoustics by exploring their potential in modeling bowed string instruments. Specifically, we investigate the bowed mass model: a one-degree-of-freedom mass-spring system driven by a nonlinear friction bow force and governed by a set of ODEs. Our findings reveal that both PINNs and PI-DeepONets effectively capture the strongly nonlinear dynamics of the bowed mass system, with PI-DeepONets demonstrating superior computational efficiency in long-time simulations. This highlights the potential of PI-DeepONets as a promising approach for advancing string instrument modeling and sound synthesis applications.
\sloppy
This study investigates the use of an unsupervised, physics-informed deep learning framework to model a one-degree-of-freedom mass-spring system subjected  to a nonlinear friction bow force and governed by a set of ordinary differential equations. Specifically, it examines the application of Physics-Informed Neural Networks (PINNs) and Physics-Informed Deep Operator Networks (PI-DeepONets). Our findings demonstrate that PINNs successfully address the problem across different bow force scenarios, while PI-DeepONets perform well under low bow forces but encounter difficulties at higher forces. Additionally, we analyze the Hessian eigenvalue density and visualize the loss landscape. Overall, the presence of large Hessian eigenvalues and sharp minima indicates highly ill-conditioned optimization. 

\sloppy
These results underscore the promise of physics-informed deep learning for nonlinear modelling in musical acoustics, while also revealing the limitations of relying solely on physics-based approaches to capture complex nonlinearities.
We demonstrate that PI-DeepONets, with their ability to generalize across varying parameters, are well-suited for sound synthesis. Furthermore, we demonstrate that the limitations of PI-DeepONets under higher forces can be mitigated by integrating observation data within a hybrid supervised-unsupervised framework. This suggests that a hybrid supervised-unsupervised DeepONets framework could be a promising direction for future practical applications.

\end{abstract}

\input{sections/introduction.tex}

\input{sections/bowedmass.tex}

\input{sections/nn}

\input{sections/results}

\input{sections/conclusion}
\nocite{*}
% \bibliographystyle{IEEEtran}
% \bibliography{DAFx25_tmpl} % requires file DAFx25_tmpl.bib

\vspace{-2mm}
{\footnotesize
\bibliographystyle{IEEEtran}
\bibliography{DAFx25_tmpl}
}

% \section{Appendix: Margin Check}
% This section shows the column margins for the text. \bigskip\newline

\end{document}

%% file: sections/introduction.tex
\section{Introduction}\label{sec:introduction}

% The physical modeling of musical instruments has been a widely explored topic, with various numerical methods employed to simulate these systems. Notable approaches include digital waveguides \cite{smith1992physical}, finite difference methods \cite{bilbao2009numerical}, and modal synthesis \cite{morrison1993mosaic}.  
In recent years,  Physics-Informed Neural Networks (PINNs) \cite{raissi2019physics}, a prominent framework in scientific machine learning (SciML), has gained significant traction in computational physics. 
The core idea of PINNs is to approximate the solution of ordinary differential equations (ODEs) or partial differential equations (PDEs) using a neural network, where the inputs are typically the space and time coordinates.
By utilizing automatic differentiation in neural networks, we can efficiently compute the required gradients in the PDE residuals, along with the losses associated with initial and boundary conditions (ICs, BCs). This results in a composite loss function with multiple competing objectives, effectively framing the problem of solving ODEs/PDEs as a multi-task deep learning challenge.
% The resulting trained models often exhibit strong generalization capabilities.
Another framework, Physics-Informed Deep Operator Networks (PI-DeepONets) was proposed to incorporate ICs as part of the input \cite{wang2021learning, wang2023long, lu2019deeponet}. This approach is effective in simulating long time-integration systems, providing an advantage over PINNs, which typically require the time-marching scheme \cite{krishnapriyan2021characterizing} or the causal training strategy \cite{wang2024respecting}.
\sloppy

Physical modelling and physics-based sound synthesis of musical instruments require solving the system's underlying governing equations \cite{bilbao2009numerical, chaigne2016acoustics}.
SciML has been utilized in musical instrument measurement methodologies, such as near-field acoustic holography, to reconstruct violin plate vibration patterns \cite{olivieri2021physics, luan2024complex, luan2025physics}. 
A PINNs approach for acoustic tube modeling was proposed in \cite{yokota2024physics}, which can be applied to model resonators of wind instruments \cite{yokota2024cnn} or the vocal tract \cite{yokota2024synthesis}, and reconstructing the acoustic field within a tube using sparsely measured pressure \cite{luan2025acoustic}, identifying physical coefficients in tubes \cite{yokota2024identification}, and determining geometric parameters of a trumpet \cite{yokota2024cnn}. 

% However, for the our best known, there is no attempt PI-DeepONets on wind instruments modeling and no for PINNs or PI-DeepONets on string instruments modeling. 

% Instead of modeling the full string dynamics, we consider a bowed mass model, representing the system as a simple oscillator. This archetypal test model is widely used in research to study numerical simulation challenges \cite{bilbao2009numerical, russo2022efficient}.

 % musical instruments has been a widely explored topic, with various numerical methods employed to simulate these systems. Notable approaches include digital waveguides \cite{smith1992physical}, finite difference methods \cite{bilbao2009numerical}, and modal synthesis \cite{morrison1993mosaic}. 
Beyond physics-informed deep learning, some studies have explored data-driven, supervised deep learning approaches for solving the governing equations in musical instrument modeling. For example, the Fourier Neural Operator has been applied to model stiff membrane vibrations \cite{de2023physical}. Recurrent neural networks, state space models and Koopman-based deep learning techniques have been used to simulate  dispersive linear lossy and nonlinear tension modulated strings \cite{schlecht2022physical, diaz2024towards}.

 However, these approaches treat the problem purely as a supervised learning task, disregarding the underlying physical knowledge encoded in the governing equations.
Physics-informed approaches are inherently more challenging than purely data-driven methods due to their unsupervised nature. As shown in \cite{gopakumar2023loss}, incorporating training data into the physics-informed framework to create a hybrid physics-informed data-driven approach can simplify the loss landscape and mitigate ill-conditioned optimization.

\sloppy
In this paper, we address the bowing simulation via PINNs and PI-DeepONets. The distinctive sound of bowed string instruments arises from the bowing mechanism, which involves continuous 
excitation resulting from bow / string interactions and nonlinear friction forces.
 When simulating bowed strings, the primary computational challenge lies in handling this nonlinearity. This challenge applies not only to traditional numerical methods, such as finite difference and finite element methods, but also to physics-informed deep learning approaches like PINNs and PI-DeepONets, as we will demonstrate in this paper. Therefore, we focus exclusively on the bowing mechanism by solving a bowed one-degree-of-freedom mass-spring system to evaluate the effectiveness of PINNs and PI-DeepONets in such simulations.
To the best of our knowledge, no existing research has explored modelling the bow-string friction interaction using deep learning, whether through data-driven or physics-informed methods. 
% We do not include experiments on data-driven or hybrid approaches, as they are generally considered less challenging compared to purely physics-informed methods.
Scenarios with different bow forces have been investigated using both PINNs and PI-DeepONets. 
The results show PINNs successfully address the problem across different bow force scenarios, while PI-DeepONets perform well under low bow forces but encounter difficulties at higher forces. Additionally, we analyze the Hessian eigenvalue density and visualize the loss landscape. The presence of large Hessian eigenvalues and sharp minima indicates highly ill-conditioned optimization.  We also demonstrate that the limitations of PI-DeepONets under higher forces can be mitigated
by integrating observation data within a hybrid supervised-unsupervised
framework. 
% Therefore, this paper specifically investigates the ill-conditioned optimization challenges associated with nonlinear bow-string friction interaction within purely physics-informed deep learning frameworks: PINNs and PI-DeepONets. 
% However, from a practical perspective, if the goal is to achieve faster and more efficient training for sound synthesis purpose, a physics-informed data-driven approach would be a promising direction. Such an approach could enhance the generalization capability of neural networks, potentially enabling real-time articulation of physical parameters.

% In this study, PINNs and PI-DeepONets are utilized to solve the nonlinear bowed one-degree-of-freedom mass-spring system. The equation-solving task is formulated as an optimization problem and carried out within the framework of physics-informed deep learning.

The remainder of this paper is organized as follows. Section~\ref{sec: bow} describes the bowed mass-spring model. Section~\ref{sec: nn} describes
PINNs and PI-DeepONets frameworks. Section~\ref{sec: result} presents the results and discussion. Finally, Section~\ref{sec: conc} summarizes the study and outlines future directions.

%% file: sections/bowedmass.tex
\section{Bowed mass-spring model}\label{sec: bow}

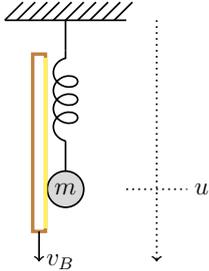
\begin{figure}[h!]
\vspace{-3mm}
    \centering
    \scalebox{0.8}{
    \begin{tikzpicture}

        % Fixed surface
        \draw[thick] (-1,5) -- (1,5);
        % Hatching lines (manual)
        \foreach \x in {-1,-0.8,...,0.8}
            \draw[thick] (\x,5) -- (\x+0.3,5.3);
            
        \draw[thick] (0,4.5) -- (0,5);
        
        % Spring (vertical)
        \draw[thick,decorate,decoration={coil,aspect=0.8,segment length=3mm,amplitude=2mm}] 
            (0,3) -- (0,4.5);

        \draw[thick] (0,2.5) -- (0,3);
        
        % Mass (circle)
        \draw[thick,fill=gray!30] (0,2.2) circle (0.3) node{\large $m$} ; 

        % block
        % \draw[ultra thick, brown] (-0.5, 1.5) -- (-0.5,3.5) -- (-0.33,3.5);
        % \draw[ultra thick, yellow!80!white] (-0.33,3.5) -- (-0.33,1.5);
        % \draw[ultra thick, brown] (-0.33,1.5) -- (-0.5,1.5);

        \draw[ultra thick, brown] (-0.45*1.2, 1.5) rectangle (-0.33,3.7*1.2); % Thick outer frame
        \draw[ultra thick, yellow!80!white] (-0.33,3.65*1.2) -- (-0.33,1.55);
    
        % velocity arrow (vertical)
        \draw[thick,->,black] (-0.435,1.5) -- (-0.435,1) node[right] {\large $v_B$};

        \draw[thick, dotted] (1,2.2) -- (2,2.2) node[right] {\large $u$} ; 
        \draw[thick, ->, dotted, black] (1.5,5) -- (1.5,1) ;

    \end{tikzpicture}
    }
    \vspace{-3mm}
    \caption{Illustration of a bowed mass-spring system. }
    \vspace{-3mm}
    \label{fig:bowed_mass}
\end{figure}

Instead of modeling the full string dynamics, we consider a bowed simple harmonic oscillator, represented by a mass-spring system with nonlinear frictional forcing. This archetypal test model is widely used in research to study numerical simulation challenges \cite{bilbao2009numerical, russo2022efficient}.
 The schematic illustration is shown in Fig. \ref{fig:bowed_mass}. The motion of the mass $m$ is described by

\begin{equation}
\begin{cases}
     u_{tt} + \omega^2 u = -  F_B \phi (\eta), \\
     \eta = u_t - v_B,
     \label{eq: bow-2}
\end{cases}
\end{equation}
where $u$ is the mass displacement~($\si{\meter}$), $\omega$ is the angular frequency~($\si{\radian\per\second}$), $F_B$ is the bow force normalized by the object's mass~($\si{\meter\per\second\squared}$), $v_B$ is the bow velocity~($\si{\meter\per\second}$), $\eta$ is the relative velocity between the mass and the bow~($\si{\meter\per\second}$), and the function $\phi(\eta)$ represents the bow friction characteristic. 
In this case, we consider the soft characteristic static friction model \cite{bilbao2009numerical}
\begin{equation}
   {\phi(\eta) = \sqrt{2a} \eta e^{-a \eta^2 + 1/2}},
   \label{eq: phi}
\end{equation}
where $a$ is a free parameter. See Fig.~\ref{eq: phi} for a plot of $\phi(\eta)$ with $a = 100$.
The derivative $\displaystyle{\frac{d\phi}{d\eta}}$ is also plotted, and the highly nonlinear region is defined as the interval between its two local minima. This region corresponds to the stick phase, characterized by low values of $\eta$, while the regions outside this interval are associated with the slip phase.
Note that this friction model is not derived from physical principles, but it provides a reasonable approximation of the discontinuity and is relatively easier to handle numerically \cite{bilbao2009numerical}.
Seeking a more accurate and physically realistic bow friction model remains an active research topic in the musical acoustics community; see \cite{matusiak2023comparison} for a comparison of different models.
However, since our primary goal is to evaluate novel computational approaches, we adopt this classical static bowed mass-spring model, which is commonly employed as a benchmark in physics-informed sound synthesis for testing numerical methods \cite{bilbao2009numerical, russo2022efficient}. Indeed, it would be worthwhile to further explore the application of the elasto-plastic friction model \cite{matusiak2023comparison}, which provides a more physically accurate representation.

% This model is more convenient for numerical implementation due to its smoothness and analytical simplicity, though it is less physically accurate—particularly in its treatment of static friction and stick-slip behavior. 

% (this model is easier to work with  numerically, however, less physically justifiable since ).

% One particular limitation of this model is that it goes to zero at phi = 0, which basically says there is no static friction force. That is quite incorrect, so I wonder if you should address that up front? Did Bilbao address that at all? The sticking phase can last for a very large portion of an oscillation!

 \begin{figure}[t!]
 \vspace{-5mm}
     \centering
     \includegraphics[width=1\linewidth]{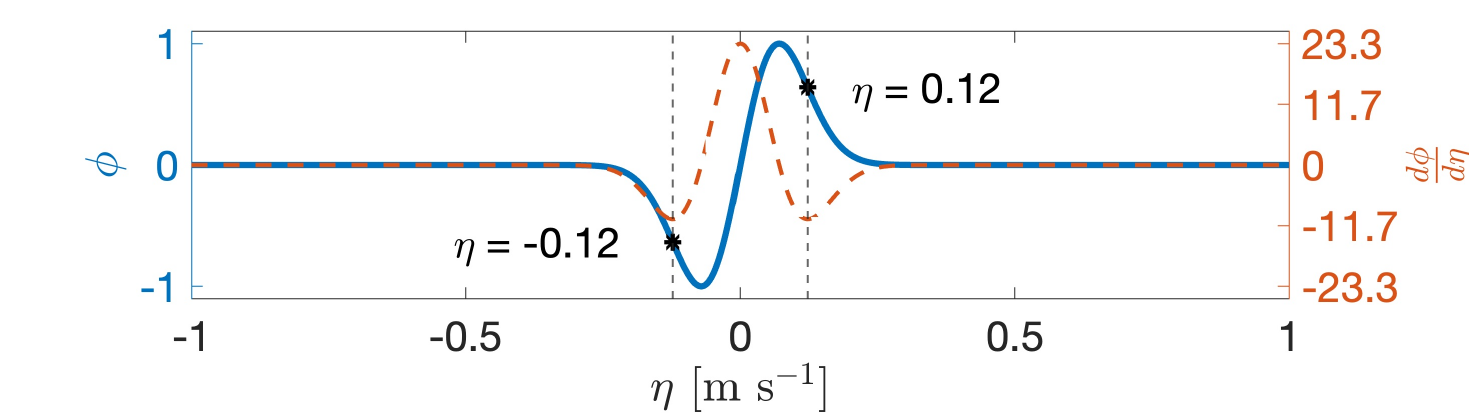}   
     \vspace{-6mm}
     \caption{Soft characteristic static friction model $\phi$ and its derivative $\displaystyle{\frac{d\phi}{d\eta}}$,  given by \eqref{eq: phi}, with $a=100$.}
     \label{fig: phi}
     \vspace{-6mm}
 \end{figure}
This one-degree-of-freedom mass-spring system is governed by two coupled ODEs \eqref{eq: bow-2}, which can be reformulated into first-order form \cite{russo2022efficient} by introducing the generalized coordinate $q$ and generalized momentum $p$, as
\begin{equation}
    q = \omega u, \quad p = u_t.
\end{equation}
Therefore, \eqref{eq: bow-2} becomes
\begin{equation}
\begin{cases}
 q_t - \omega p = 0, \\
    p_t + \omega q + F_B \phi (\eta) =0.
\end{cases}
\label{eq: bow-1}
\end{equation}
In the following, we use this first-order form \eqref{eq: bow-1} for the simulation.
To solve \eqref{eq: bow-1}, we require two initial conditions (ICs) for both $p|_{t=0}, q|_{t=0}$.

%% file: sections/nn.tex
\section{Neural Networks}\label{sec: nn}
% Both PINNs and PI-DeepONets are employed to solve the ODEs \eqref{eq: bow-1}.
We present the solution of the bowed mass-spring system by formulating it as an optimization problem, which is then solved using the PINNs and PI-DeepONets frameworks.

\subsection{Modified FCNN}
\begin{figure*}
\vspace{-4mm}
    \centering
\includegraphics[width=0.85\linewidth]{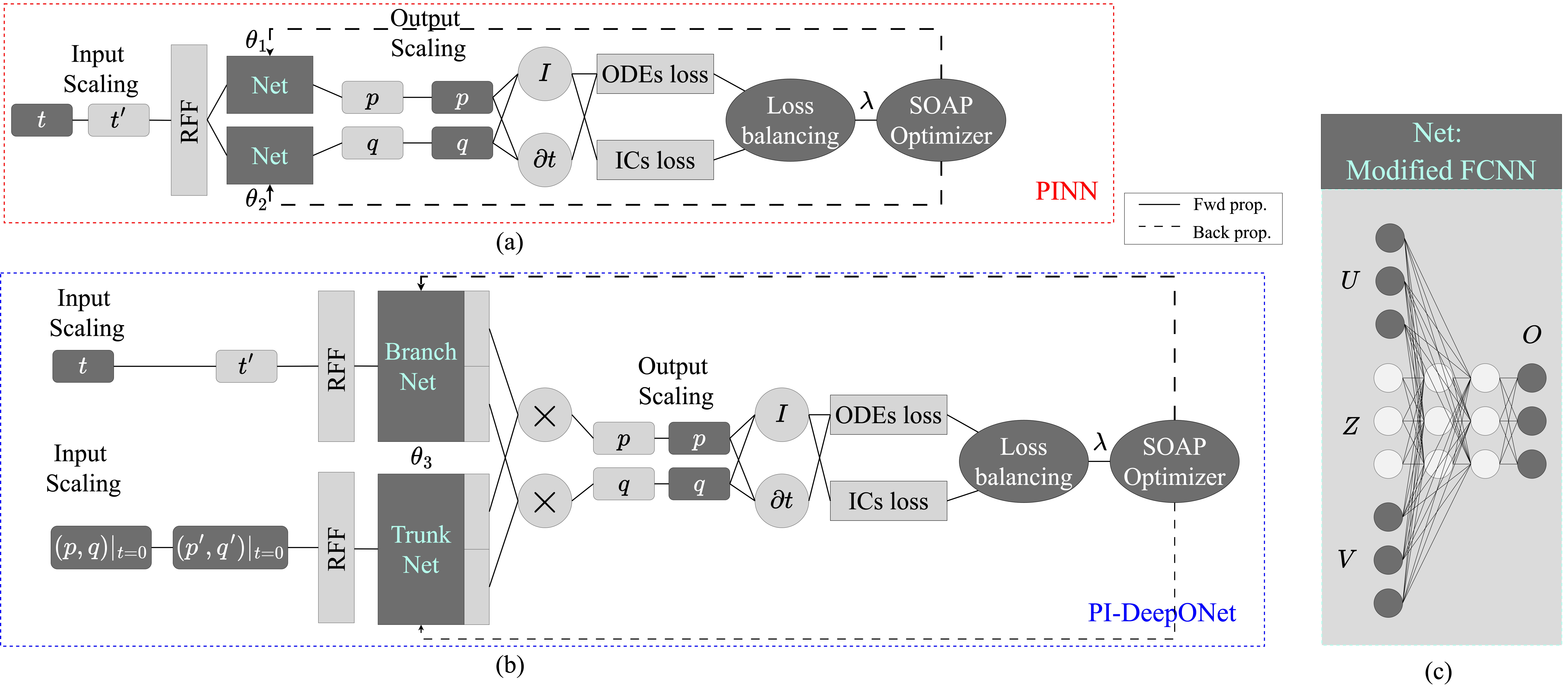}
\vspace{-3mm}
    \caption{Network Architectures: (a) PINNs, (b) PI-DeepONets, and (c) Modified FCNN \cite{wang2021understanding}.  Note that (c) serves as a component of the PINNs' Nets, as well as the Branch and Trunk Nets in PI-DeepONets. The legend indicating forward propagation (fwd prop.) and backpropagation (back prop.) applies to both (a) and (b).}
   \vspace{-5mm}
    \label{fig:nn arch}
\end{figure*}

A modified fully-connected neural network (FCNN) \cite{wang2021understanding}, inspired by attention mechanisms, is used in this study. It outperforms standard FCNNs in PINNs by capturing multiplicative interactions between input dimensions and including residual connections \cite{wang2021understanding}.
Given the network inputs $X$  and outputs $O$, the forward pass is defined by propagating $X$ through the network layers to compute $O$, through \cite{wang2021understanding}

\begin{equation}
    \begin{aligned}
       & U = \sigma(XW^U + b^U), \quad V = \sigma (XW^V + b^V), \\&
        H^{(1)} = \sigma(XW^{Z,1}+ b^{Z,1}), \\&
        Z^{(k)} = \sigma(H^{(k)} W^{Z,k}+ b^{Z,k}), \quad k=1,...,L, \\&
         H^{(k+1)} = (1-Z^{(k)}) \odot U + Z^{(k)} \odot V, \quad k=1,...,L, \\&
         O = H^{(L+1)} W^{O} + b^{O},
    \end{aligned}
\end{equation}
where $\sigma$ is the activation function and $\odot$ denotes element-wise multiplication. The network parameters are
\begin{equation}
    \theta = \{W^U, b^U, W^V, b^V, (W^{Z,k}, b^{Z,k})_{k=1}^{L}, W^O, b^O \},
\end{equation}
where $W^{(\cdot)}$
  and $b^{(\cdot)}$
  denote the weight matrices and bias vectors of the corresponding layer. 
The channel or layer sizes of $U, V, Z^{(k)}$ and $O$ are denoted as $c_U, c_V, c_{Z^{(k)}}$ and $c_O$, respectively. $Z$ has $L$ layers in total.
A graphical representation of this modified FCNN is shown in Fig. \ref{fig:nn arch} (c).

\subsection{PINNs} \label{sec: pinns}
PINNs are formulated as
\begin{equation}
\begin{cases}
        p = \mathcal{F}_{1, \theta_1} (t),  \\
    q = \mathcal{F}_{2, \theta_2} (t),
\end{cases}
\end{equation}
where  $\mathcal{F}_{1, \theta_1}$ and $\mathcal{F}_{2, \theta_2}$ represent two distinct networks with identical architecture.
Since  $\mathcal{F}_{1, \theta_1}$ and $\mathcal{F}_{2, \theta_2}$ follow the same flow, we illustrate the process using one as an example.
The input to  $\mathcal{F}_{1, \theta_1}$ is the time coordinate $ t \in [0, t_N] \subset \mathbb{R}^{1 \times N}$, a single channel vector sampled at $N$ points, which is first scaled by scaling factor $s^t$ for normalization. Then the Random Fourier Feature (RFF) embedding \cite{tancik2020fourier} is employed with a scale parameter $\sigma'$, which controls the range of frequencies in the embedding, and is applied with an encoding size of $c_{RFF}$. RFF maps input data into a higher-dimensional space using sinusoidal transformations, helping to mitigate \textit{spectral bias} phenomenon—the tendency of neural networks to learn low-frequency components faster while struggling with high-frequency ones \cite{rahaman2019spectral}. RFF has proven effective in overcoming spectral bias in PINNs \cite{wang2021eigenvector}.
The embedded features next serve as the input to the modified FCNN. The output of the FCNN is $p \subset \mathbb{R}^{1 \times N}$, which are scaled by scaling factor $s^{p}$ ($s^q$ for the other case).
Automatic differentiation is then used to compute the loss functions, which encompass the PDE and IC  losses, all formulated as mean squared error (MSE) terms. 
Predicted values are denoted with a hat $\hat{\cdot}$.
The PDE losses are expressed as
\begin{equation}
\begin{cases}
       \displaystyle{ \mathcal{L}_{ODE_1} = \frac{1}{N_{ODE}}  \bigg \| \hat{q}_t - \omega \hat{p}  \bigg \|,  }\\ \\
        \displaystyle{ \mathcal{L}_{ODE_2} = \frac{1}{N_{ODE}} \bigg  \| \hat{p}_t + \omega \hat{q} + F_B  \sqrt{2a} \eta e^{-a \eta^2 + 1/2} \bigg  \| }, \\  \qquad \qquad \qquad \qquad  \qquad \qquad  \qquad \quad t \in [0,t_N].
\end{cases}
\end{equation}
The IC losses are 
\begin{equation}
\begin{cases}
 \displaystyle{ \mathcal{L}_{IC_1} = \frac{1}{N_{IC}}  \bigg \| \hat{p}|_{t=0} - p|_{t=0} \bigg \|,  }\\ \\
  \displaystyle{ \mathcal{L}_{IC_2} = \frac{1}{N_{IC}}  \bigg \| \hat{q}|_{t=0} - q|_{t=0} \bigg \|. } 
\end{cases}
\end{equation}
$N_{ODE}$ and $N_{IC}$ represent the respective numbers of collocation points used for the ODE and IC loss computations. 
Then the total loss function is 
\begin{equation}
\begin{aligned}
    \mathcal{L} = \lambda_{ODE_1} \mathcal{L}_{ODE_1} +  \lambda_{ODE_2} \mathcal{L}_{ODE_2} + \\ \lambda_{IC_1} \mathcal{L}_{IC_1} + \lambda_{IC_2} \mathcal{L}_{IC_2},
\end{aligned}
\label{eq: tot loss}
\end{equation}
with $\lambda_{ODE_1} $, $\lambda_{ODE_2}$,  $ \lambda_{IC_1}$ and $ \lambda_{IC_2}$ as the loss function weights. We employ learning rate annealing for loss balancing to determine these weights, as described in \cite{wang2021understanding}. 
The total loss function is then fed into the optimizer, where a gradient descent routine is applied via back propagation to update $\theta_1$  and $\theta_2$. 
A diagram for the architecture of PINNs is shown in Fig. \ref{fig:nn arch} (a).

However, PINNs often encounter {failure mode} challenges when the optimization problem is ill-conditioned \cite{krishnapriyan2021characterizing}. Various approaches have been proposed to address these issues, and a comprehensive review of PINNs training strategies can be found in \cite{wang2023expert}.  In long-time integration problems (always the case for sound synthesis or musical acoustic simulation), it is extremely difficult to solve the entire time domain simultaneously. Additionally, studies have shown that continuous-time PINNs models can violate causality, making them prone to converging toward incorrect solutions \cite{wang2024respecting}.
To overcome the time related issues, we utilize the \textit{time-marching} scheme \cite{krishnapriyan2021characterizing} and \textit{causal training} strategy \cite{wang2024respecting}.
For the time-marching scheme, the entire time domain is segmented into $M_{tm}$ subdomains (or windows), as 
\begin{equation}
\begin{aligned}
        t^{(i)} \in [t_i, t_{i+1}] \subset \mathbb{R}^{1 \times N_i}, \text{for } i= 0, ..., M_{tm}-1, \\ \text{where } \sum_{i=0}^{M_{tm}-1}N_i =N, t_0 =0, t_{M_{tm}} = t_N.
\end{aligned}
\end{equation}
In other words, a separate network is utilized and trained after the optimization of the previous subdomain is solved, following a time-marching routine and ultimately resulting in a total of $M_{tm}$ sub-networks.
The schematic of time-marching for PINNs is presented in Fig.~ \ref{fig: domain}. 
For a more challenging task, we adopt the causal training strategy, for which each subdomain $t^{(i)}$ is divided into $M_{cas}$ chunks
\begin{equation}
\begin{aligned}
        t^{(j)} \in [t_j, t_{j+1}] \subset \mathbb{R}^{1 \times N_j}, \text{for } j= 0, ..., M_{cau}-1, \\ \text{where } \sum_{j=0}^{M_{cau}-1}N_j =N_i, t_0 =t_i, t_{M_{cau}} = t_{i+1}.
\end{aligned}
\end{equation}
The training of the $i$-th sub-network starts with the time samples from $t^{(j=1)}$. Once the loss condition $\mathcal{L}_{ODE_1} < \eta_{cau}$ is satisfied, the time samples are augmented by adding samples from $t^{(j=2)}$, continuing this process iteratively until the last chunk is included. Note that the data augmentation procedure occurs within a single network training process, making it fundamentally different from time-marching.  A schematic view of time-marching and causal training is provided in Fig.~\ref{fig: domain}.

% \textcolor{red}{soap: preconditoning frequency}

\vspace{-2mm}
\subsection{PI-DeepONets} 
PI-DeepONets \cite{wang2023long} are formulated as
\begin{equation}
        (p, q) = \mathcal{F}_{3, \theta_3} (t, (p,q)|_{t=0}).
\end{equation}
The inputs to  $\mathcal{F}_{3, \theta_3}$ are  
 $ t \in [0, t_M] \subset \mathbb{R}^{1 \times M}$,  similar to PINNs, scaled by $s_i^t$ and  $(p,q)|_{t=0} \in [p_{min}, p_{max}] \times [q_{min}, q_{max}]  \subset \mathbb{R}^{2 \times M}$, scaled by $s^{p,q}$. 
Then, the RFF embedding is applied to both  $t$  and $(p,q)|_{t=0} $, which are then fed into the branch  and trunk net, respectively. 
Both networks are modified FCNNs. The latent features from both the branch and trunk network outputs are split into two equal parts, with each half merged via a dot product to separately produce the outputs $p \subset \mathbb{R}^{1 \times M}$ and $q \subset \mathbb{R}^{1 \times M}$, which are scaled by $s^{p,q}$. Then, the backpropagation process follows a similar routine as PINNs. A diagram for the architecture of PI-DeepONets is shown in Fig.~\ref{fig:nn arch} (b).

\subsection{PINNs vs. PI-DeepONets}
\begin{figure}
\vspace{-1mm}
    \centering
\includegraphics[width=1\linewidth]{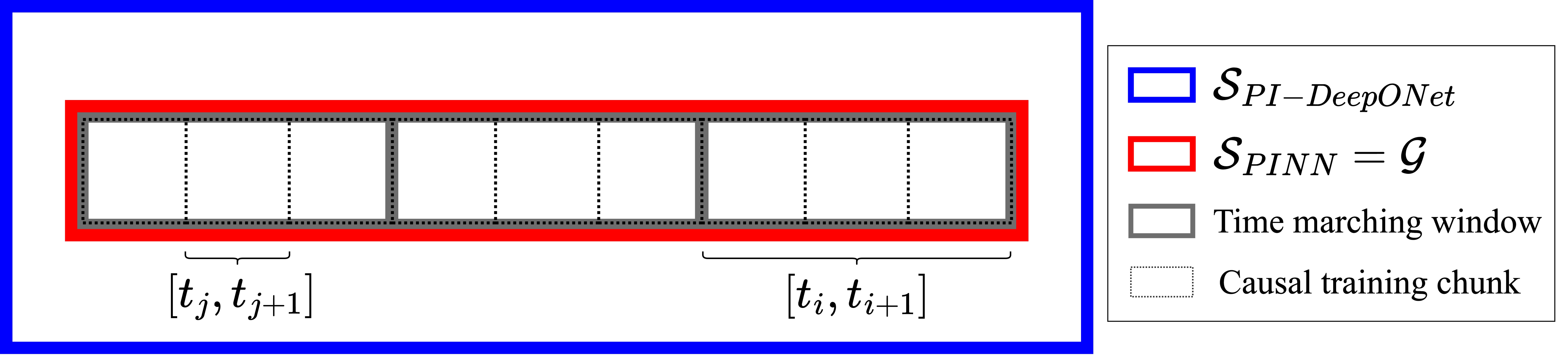}
 \vspace{-2mm}  
    \caption{The solution spaces that PINNs and PI-DeepONets aim to solve.}
    \label{fig: domain}
      \vspace{-6mm}  
\end{figure}

It is natural to ask about the differences between PINNs and PI-DeepONets. 
From the input perspective, PINNs use only the coordinate $t$ as input. In contrast, PI-DeepONets take $t$ as input to the branch network and also incorporate the initial condition $(p,q)|_{t=0}$ as input to the trunk network.
From the solution perspective, 
the ODEs themselves \eqref{eq: bow-1} define an infinite domain $\mathcal{S}$ (with $ t \in [0,\infty] $), but the solution space of interest—referred to as the goal space $\mathcal{G}$—is a subspace of this infinite domain, with $\mathcal{G} \subset \mathcal{S}$. In the case of PINNs, solving a long-time integration problem requires employing multiple ($M_{tm}$) neural networks over different time windows to cover $\mathcal{G}$. Therefore, under the ideal assumption that PINNs can perfectly solve the ODEs, we have
\begin{equation}
    \mathcal{S}_{PINN} = \bigcup_{i=0}^{M_{tm}-1} \mathcal{S}_{PINN}^{(i)}, \quad
    \mathcal{G} = \mathcal{S}_{PINN}.
\end{equation}
However, PI-DeepONets, by randomly sampling ICs  $p(0) $ and $q(0)$, inherently capture $\mathcal{G}$ within their solution space $\mathcal{S}_{PI-DeepONet}$. Under the ideal assumption that PI-DeepONets can perfectly solve the ODEs, we obtain 
\begin{equation}
    \mathcal{G} \subset \mathcal{S}_{PI-DeepONet}.
\end{equation}
Essentially, the solutions corresponding to these randomly sampled ICs do not necessarily lie within $\mathcal{G}$.
In other words, PI-DeepONets operate over a significantly larger solution space compared to PINNs. Ideally, if both methods achieve equally good results, PI-DeepONets might be the preferred choice.  However, given the broader solution space, the complexity of PI-DeepONets is inherently higher than that of PINNs. A visualization of the solution space is provided in Fig.~\ref{fig: domain}.
% However, we will see in the following section, when solving ODEs formulated in neural networks optimization problem, when the problem itself is ill-conditioned, it is hard for both PINNs and PI-DeepONets to solve. while  

%% file: sections/results.tex
\vspace{-2mm}
\section{Simulation Results}\label{sec: result}
\vspace{-3mm}

\begin{table*}[h!]
\vspace{-2mm}
 \captionsetup{skip=0pt} 
 \vspace{-4mm}
 \caption{Hyperparameters and training strategies for PINNs and PI-DeepONets.}
 \begin{center}
 \begin{tabular}{l@{\hspace{3pt}}|l@{\hspace{3pt}}|l@{\hspace{3pt}}| l@{\hspace{3pt}}|l@{\hspace{3pt}} l@{\hspace{3pt}} l@{\hspace{3pt}}|l@{\hspace{3pt}} l@{\hspace{3pt}}|l@{\hspace{3pt}} l@{\hspace{3pt}}|l@{\hspace{3pt}}|l@{\hspace{3pt}}  }
  % \hline
  Net & $F_B$ & $s^t$ &  $s^{p,q}$ & cas. & $M_{cau}$ &$\eta_{cau}$ & tm.& $M_{tm}$ & RFF & $\sigma'$ & ann.& runtime [\si{\hour}]\\ \hline
  PINNs &  10 & 0.1 & 0.2 & $\times$ &- &- &$\checkmark
  $ & 3 &$\checkmark$ & 1 & $\times $ & 38.76 \\
  PINNs &  100 & 0.03 & 0.2 & $\times$ & - & - &$\checkmark$ & 3 & $\checkmark$ & 1 & $\times $ &38.14\\
  PINNs &  1000 & 0.01 & 1 &$\checkmark$ & 50  & 0.1 &$\checkmark$ & 5 & $\checkmark$ & 3 & $\checkmark$  & 14.59\\
  PI-DeepONets &  10 & 0.01 & 0.35 &$\times$ & - &- & $\times$ & - & $\checkmark$& 1 & $\checkmark$ & 36.62\\
  PI-DeepONets &  100 & 0.01 & 0.35 & $\times$ & - &- & $\times$ & - & $\checkmark$& 1 & $\checkmark$ & 21.41\\
  PI-DeepONets & 1000 & 0.01 & 2 &$\times$ & - &- & $\times$ & - & $\checkmark$& 3 & $\checkmark$ & 14.22\\
  % \hline
 \end{tabular}
\end{center}
  \vspace{-10mm}  
 \label{tab: stra}
\end{table*}

\subsection{Implementation}

We validate the PINNs and PI-DeepONets for $\omega = 2 \pi f, f= $ \SI{100}{\hertz}, $a= 100$, and $v_B = $ \SI{0.2}{\meter \per \second}, considering three different values of $F_B$ set as $10$, $100$, and $1000$. 
The choice of varying $F_B$ is motivated by its influence on the bowing mechanism, resulting in different waveforms.
% \textcolor{red}{Additionally, an example with $f=$ \SI{1}{\hertz}, $F_B =100$ and $v_B = $ \SI{0.2}{\meter \per \second} is presented, demonstrating a case where RFF is unnecessary.}
For both PINNs and PI-DeepONets, the hyperbolic tangent ($\tanh$) activation functions are utilized, the RFF encoding size, $c_{RFF}$, is set to $50$ and the resulting layer $U$ and $V$ have $c_U = c_V=100$ channels. 
For PINNs, each modified FCNN consists of $L=4$ $Z$ layers, with each layer containing $c_{Z^{(k)}}=100$ channels. The output layer $O$ has $c_O=1$ channel.
For PI-DeepONets, each modified FCNN in both the branch and trunk net consists of $L=6$ $Z$  layers, with each layer containing $c_{Z^{(k)}}=100$ channels. The output layer $O$ has $c_O=200$ channels.

For PINNs, $N_{{ODE}_1} =N_{{ODE}_2} = 1000$, $N_{{IC}_1}= N_{{IC}_2}= 1$. The ODEs input $t$ are uniformly sampled within the interval $[0, s^t]$. The training of PINNs follows a full-batch paradigm.
For PI-DeepONets, $N_{{ODE}_1} =N_{{ODE}_2} = 10000 \times 1000$, $N_{{IC}_1}= N_{{IC}_2}= 10000$.
To construct the dataset, we first generate a single group of data. Each group starts with an IC input $(p,q)$ randomly sampled from the range $[-s^{p,q}, s^{p,q}]$ at $t=0$ (we set $s^p = s^q = s^{p,q}$ for all cases). The corresponding ODEs inputs $(p,q)$ are then repeated $1000 $ times while $1000$ values of $t$ are randomly sampled from $[0, s^t]$.
This process is repeated $10000$ times to create the final input dataset. 
PI-DeepOnets employ a mini-batch training strategy with a batch size of $50000$.
We employ the state-of-the-art second order optimizer ShampoO with Adam in the Preconditioner's eigenbasis (SOAP) \cite{vyas2024soap} for both PINNs and PI-DeepONets. 
SOAP has been demonstrated to efficiently approximate the Hessian preconditioner, leading to significant performance improvements in PINNs \cite{wang2025gradient}.
Note that either full-batch training or a large batch size is used, as it is suggested in \cite{vyas2024soap} that a large batch size enhances the performance of the SOAP optimizer. 
The other hyperparameters for SOAP remain at their default values.  
The initial learning rate is set to $0.003$.  
An exponential learning rate scheduler is applied with a decay rate of $0.9$.  
The decay step is set to $10000$ for PINNs and $3000$ for PI-DeepONets.
% The value of $s^t$ and $s^{p,q}$ are shown in Table~\ref{tab: stra}.
A loss weight balance strategy, learning rate annealing (ann.) \cite{wang2021understanding}, may be employed for some cases.
If the adaptive loss weight annealing strategy is not applied, weights are manually set as $\lambda_{{ODE}_1} =\lambda_{{ODE}_2} = 10, \lambda_{{IC}_1} =\lambda_{{IC}_2} = 1 \times 10^6 $.
% The neural networks are initialized with default setting.
The training strategies and other hyperparameters are summarized in Table~\ref{tab: stra}.
It is worth mentioning that different hyperparameters may be used for different values of 
$F_B$, depending on the complexity of the optimization task. As discussed in Section~\ref{sec: pinns}, incorporating the ODEs into the loss function can lead to an ill-conditioned optimization problem, which may cause the network to encounter failure modes \cite{krishnapriyan2021characterizing}.
Intuitively, more difficult tasks demand a more careful design, potentially incorporating additional strategies, more time-marching windows, and more chunks for causal training, when such techniques are used. In other words, these hyperparameters are sensitive to $F_B$.
The training processes are terminated once the monitored loss converges.
Notice that to successfully train PINNs for $F_B=1000$, the causal training process may terminate early at an intermediate chunk rather than progressing to the final chunk if incorporating data from the next chunk introduces a bias that compromises the accuracy of the entire window.
The implementation is carried out using PyTorch and an NVIDIA GeForce RTX 4080 GPU with 16 GB VRAM. The code and accompanying sound samples are available on github\footnote{\href{https://github.com/Xinmeng-Luan/bowmass-dafx}{https://github.com/Xinmeng-Luan/bowmass-dafx}}.

\begin{figure*}
\vspace{-9mm}
    \centering
    \subfloat{\includegraphics[width=0.3\linewidth]{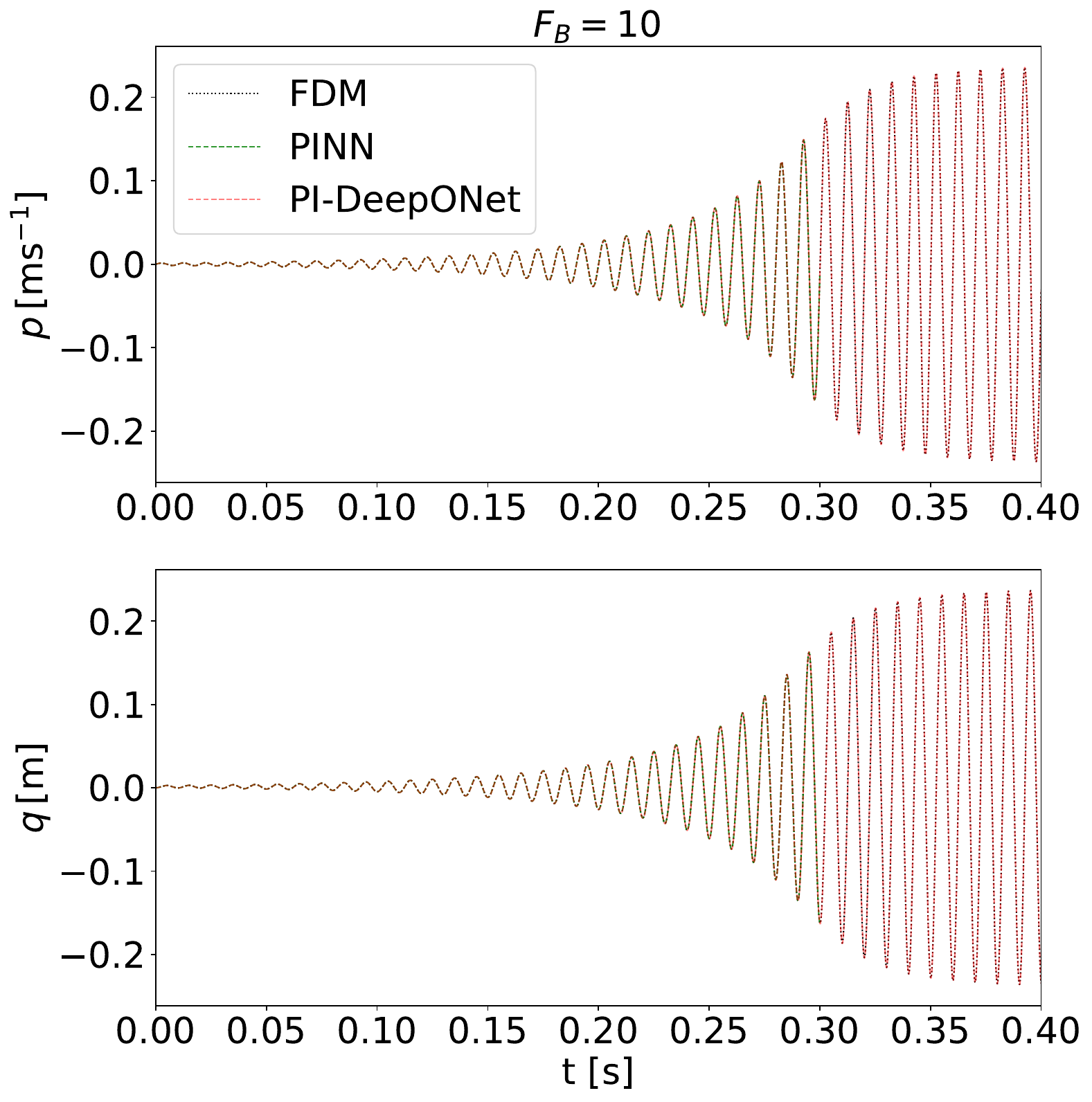}}
    \subfloat{\includegraphics[width=0.3\linewidth]{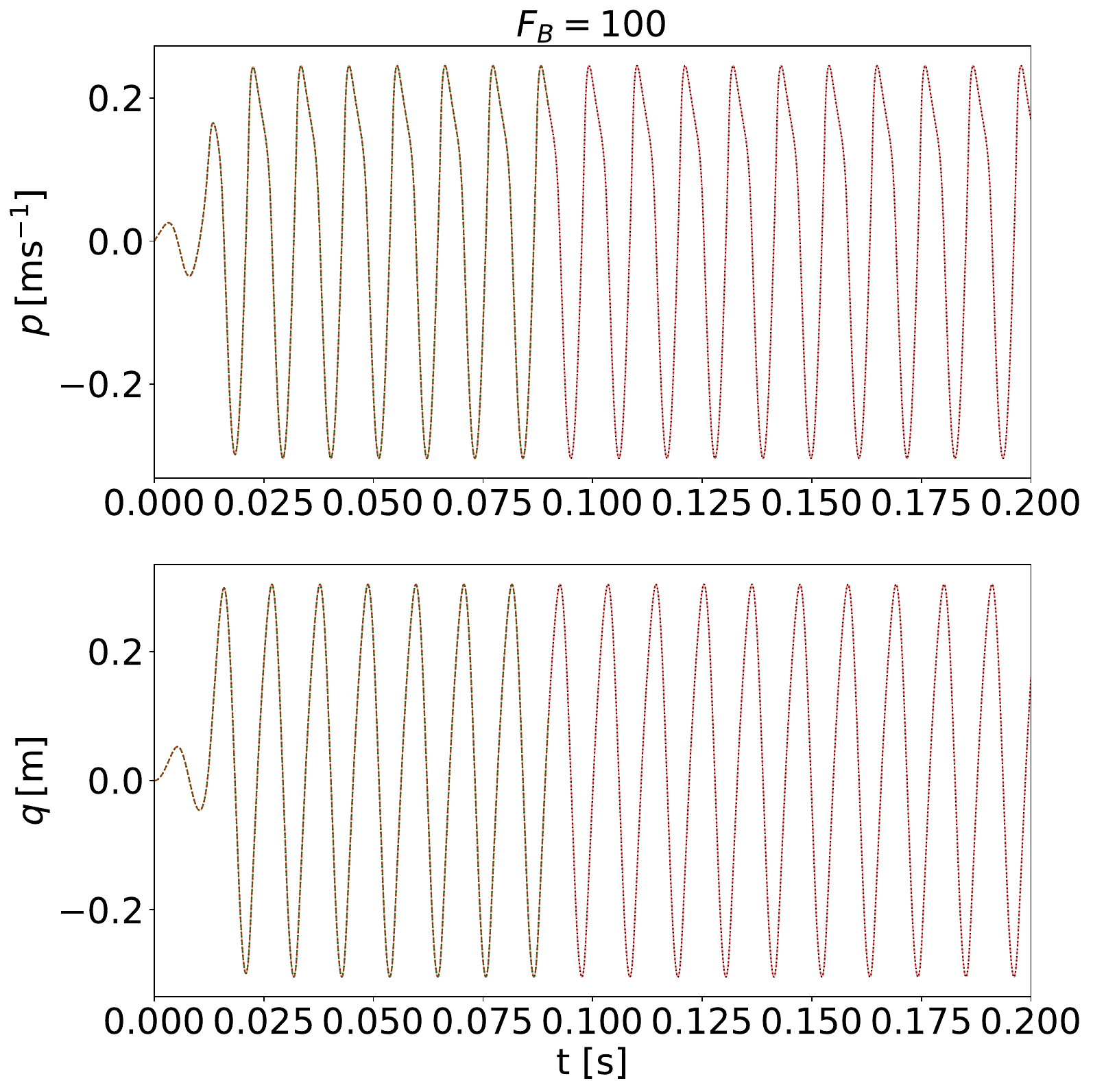}}
    \subfloat{\includegraphics[width=0.3\linewidth]{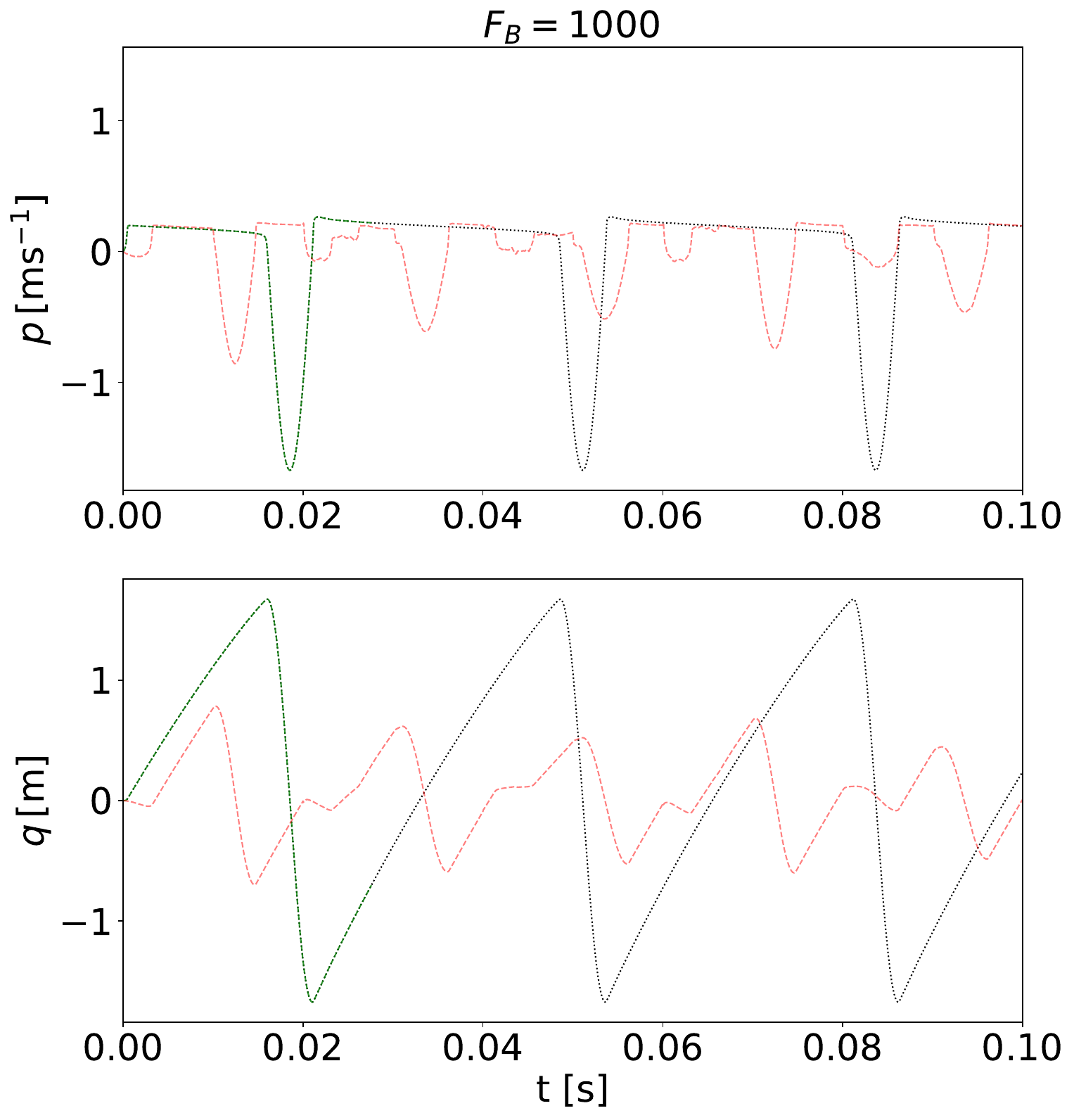}}
    
    \vspace{-4mm} 
    \subfloat{\includegraphics[width=0.3\linewidth]{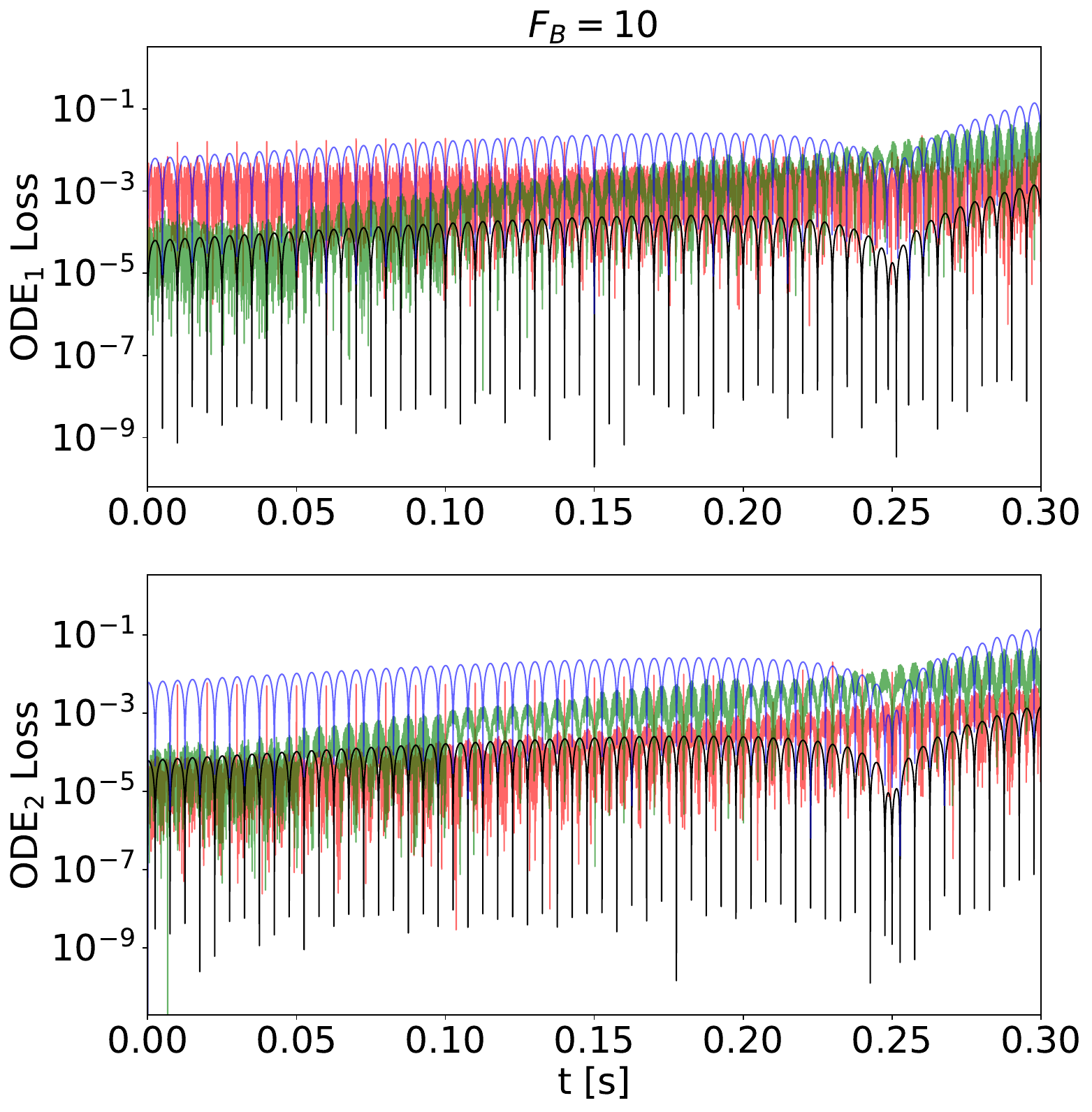}}
    \subfloat{\includegraphics[width=0.3\linewidth]{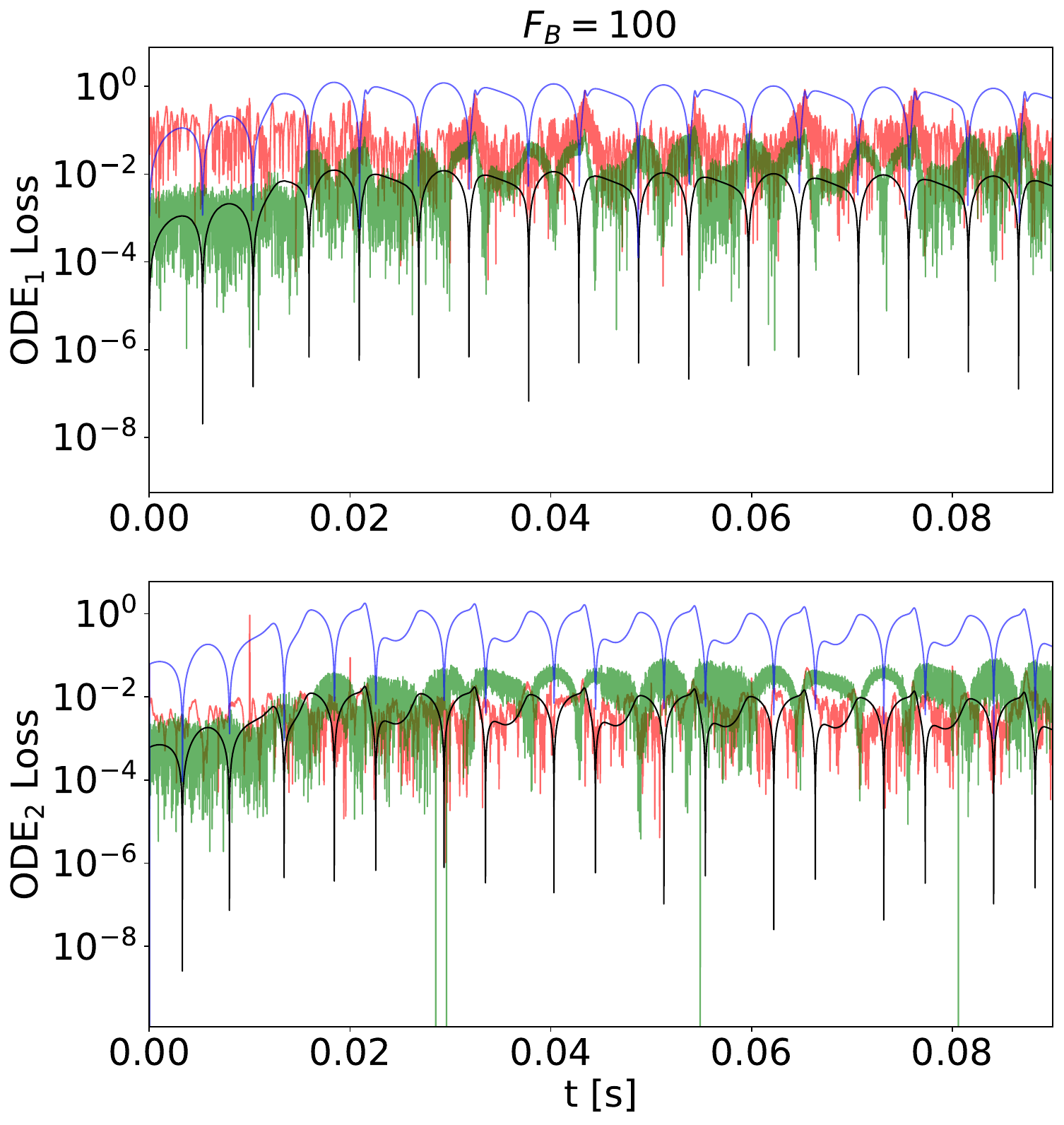}}
    \subfloat{\includegraphics[width=0.3\linewidth]{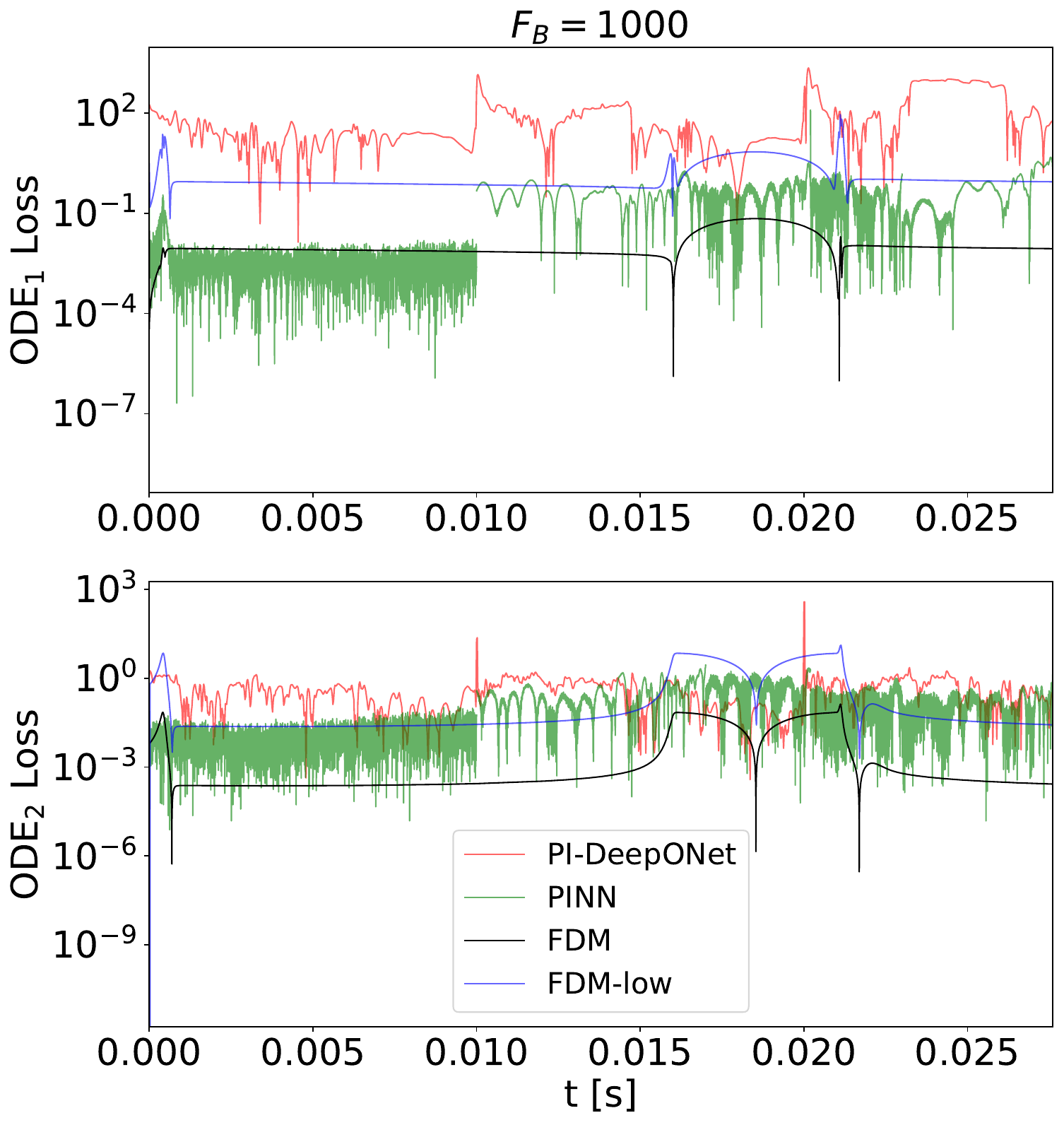}}
    
    \vspace{-4mm} 
    \subfloat{\includegraphics[width=0.3\linewidth]{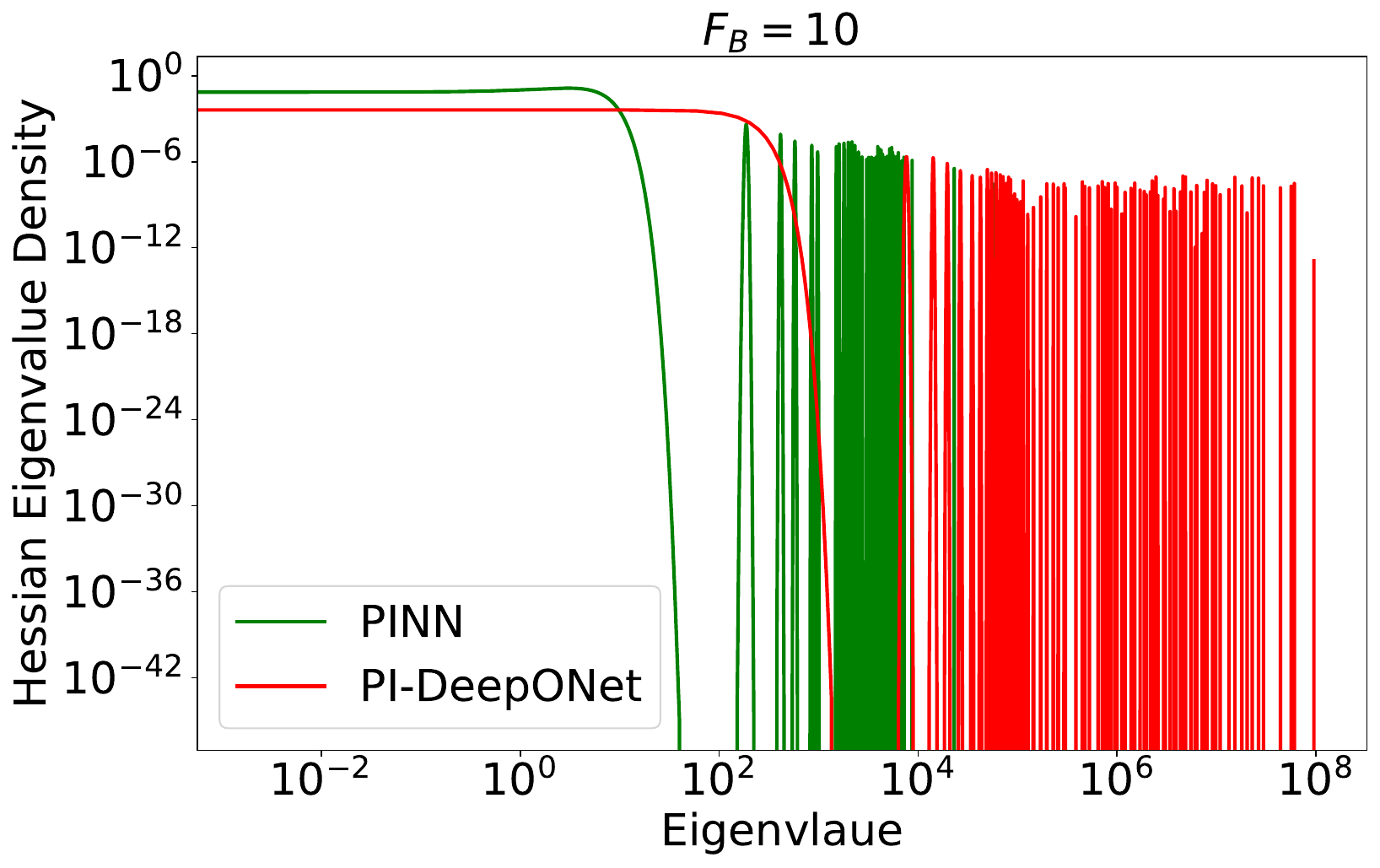}}
    \subfloat{\includegraphics[width=0.3\linewidth]{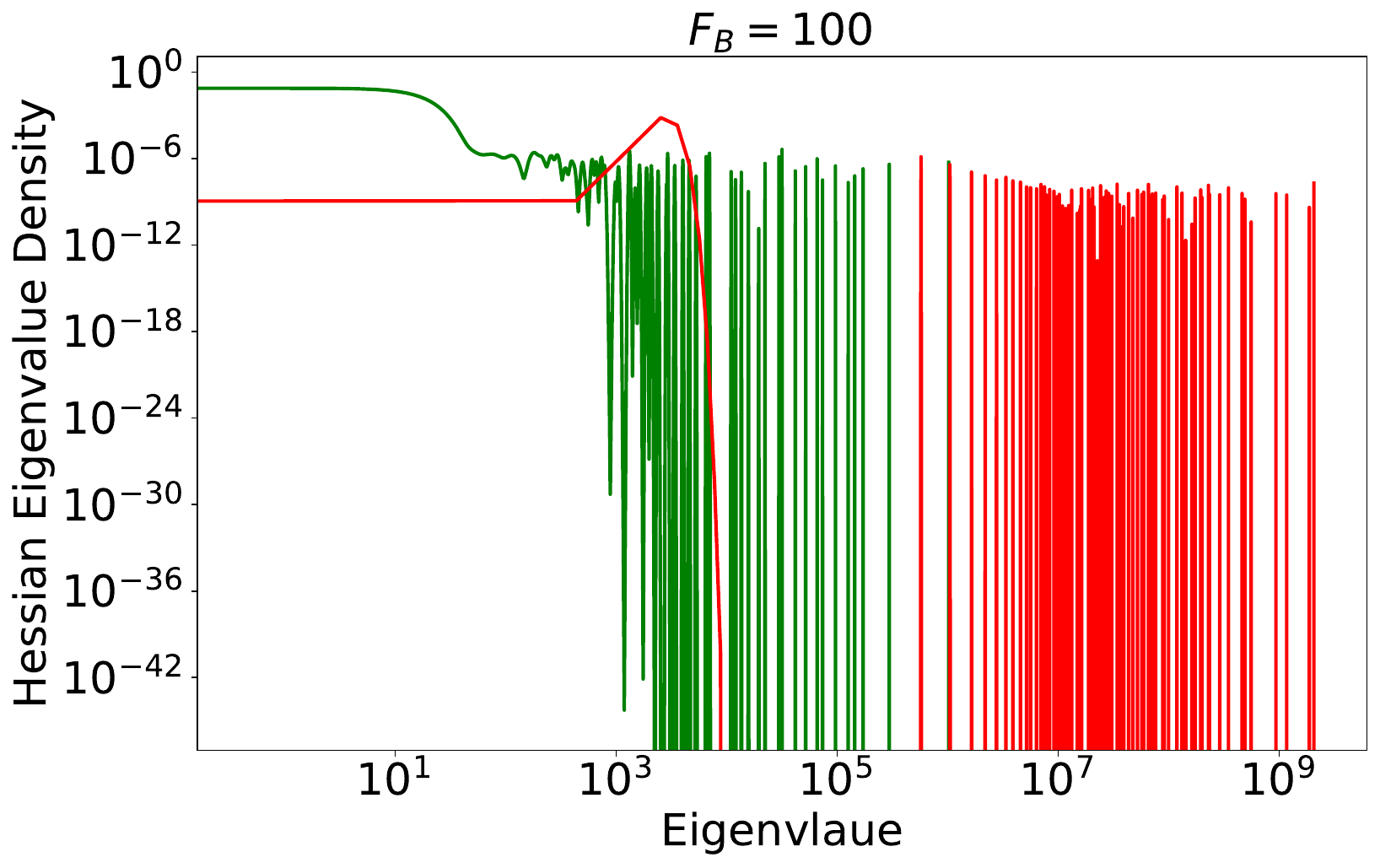}}
    \subfloat{\includegraphics[width=0.3\linewidth]{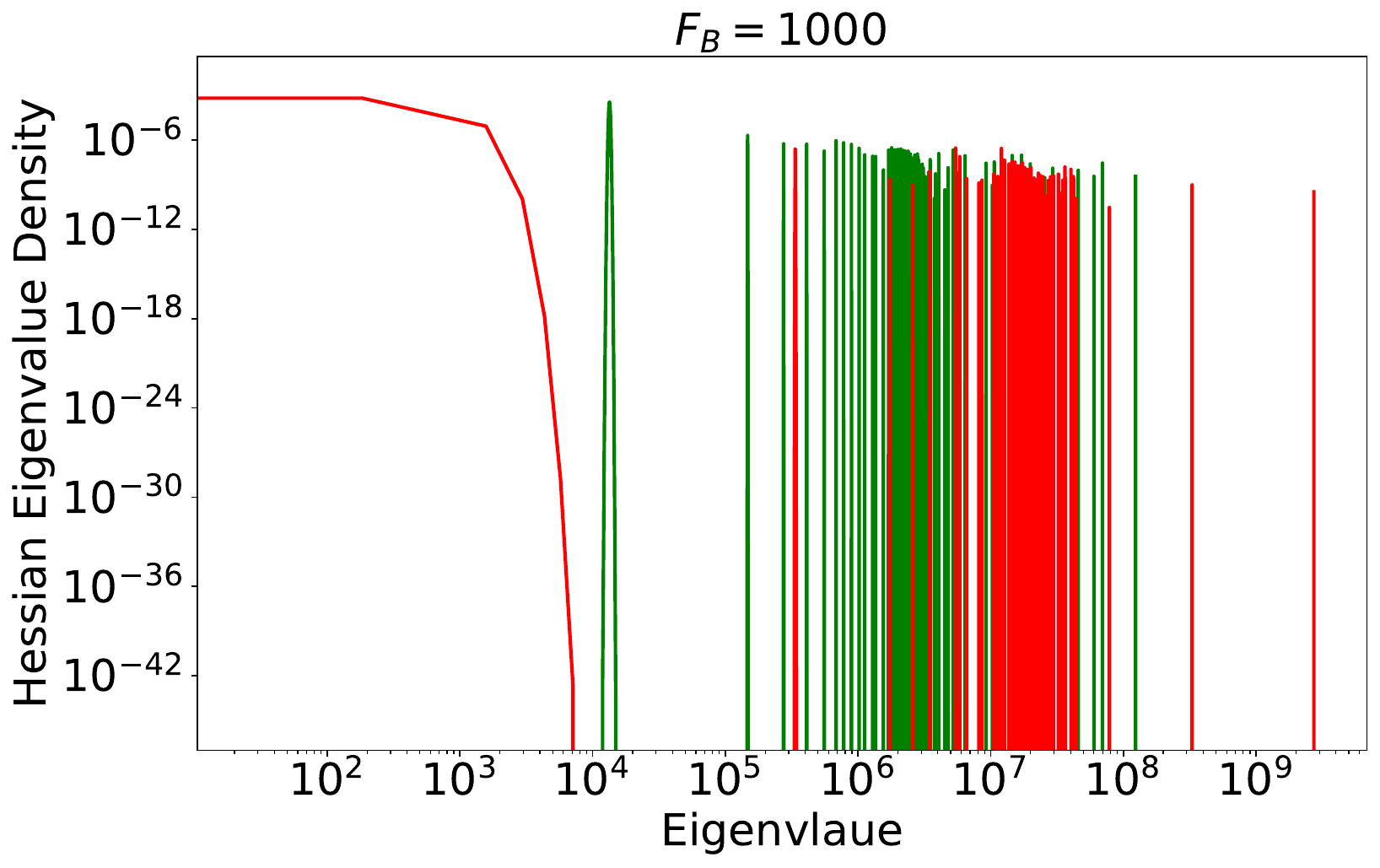}}

    \vspace{-4mm} \subfloat{\includegraphics[width=0.2\linewidth] {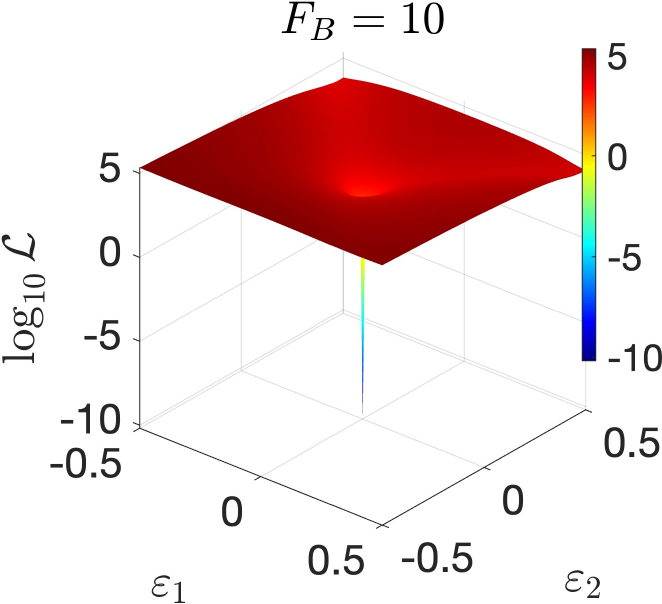}}
    \hspace{0.1\linewidth} \subfloat{\includegraphics[width=0.2\linewidth]{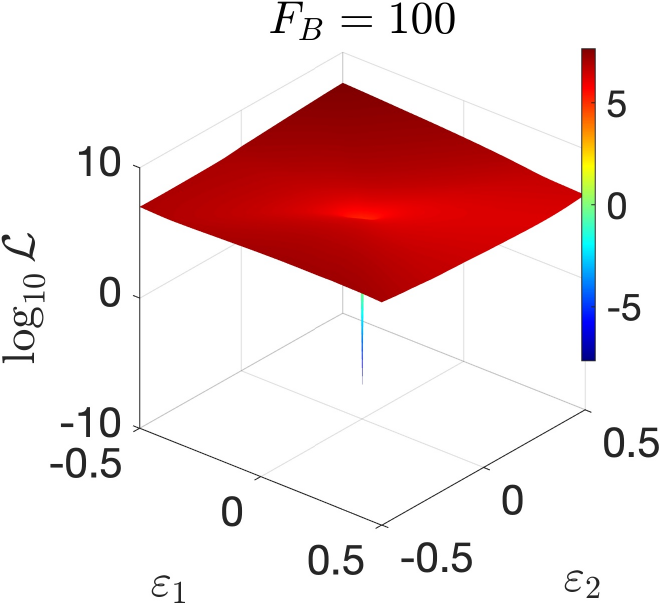}}
    \hspace{0.1\linewidth} \subfloat{\includegraphics[width=0.2\linewidth]{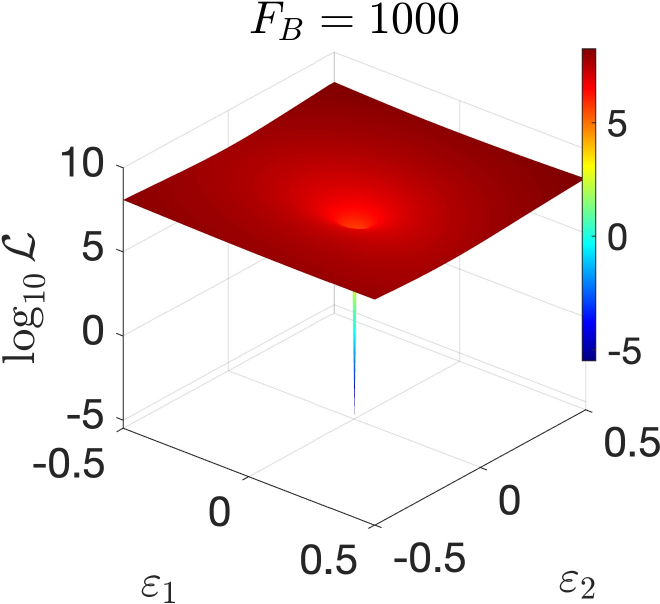}}

\vspace{-4mm} 
    \subfloat{\includegraphics[width=0.2\linewidth]{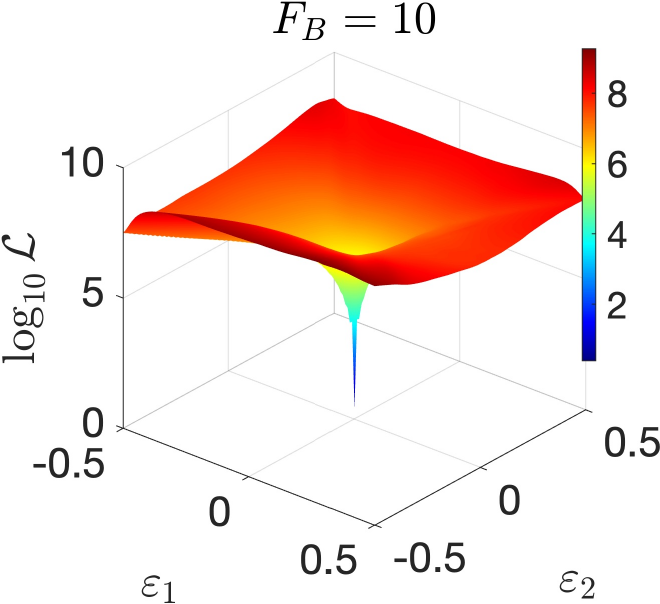}}
    \hspace{0.1\linewidth} \subfloat{\includegraphics[width=0.2\linewidth]{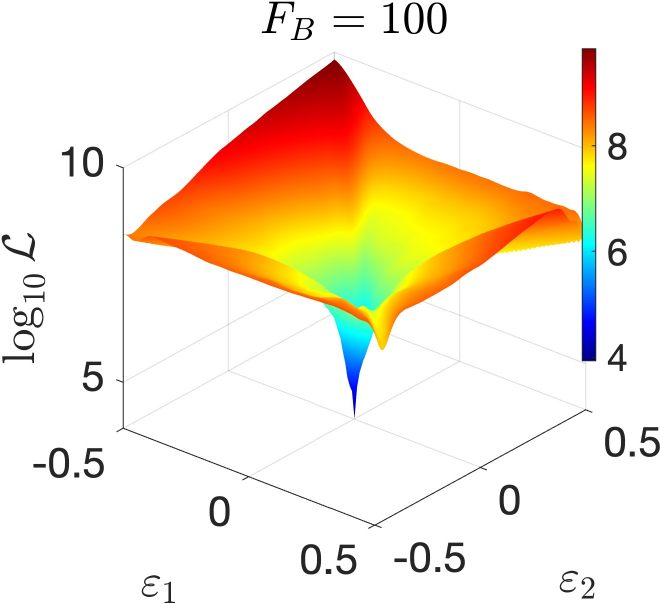}}
    \hspace{0.1\linewidth} \subfloat{\includegraphics[width=0.2\linewidth]{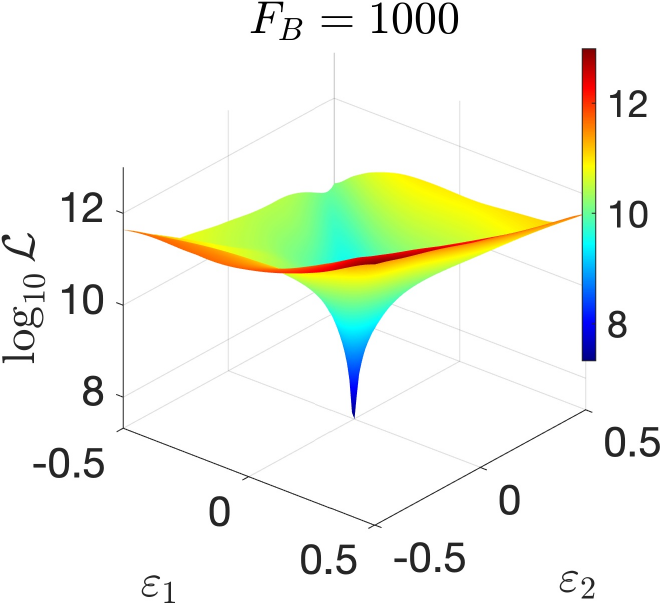}}

\vspace{-4mm}
    \subfloat{\includegraphics[width=0.3\linewidth]{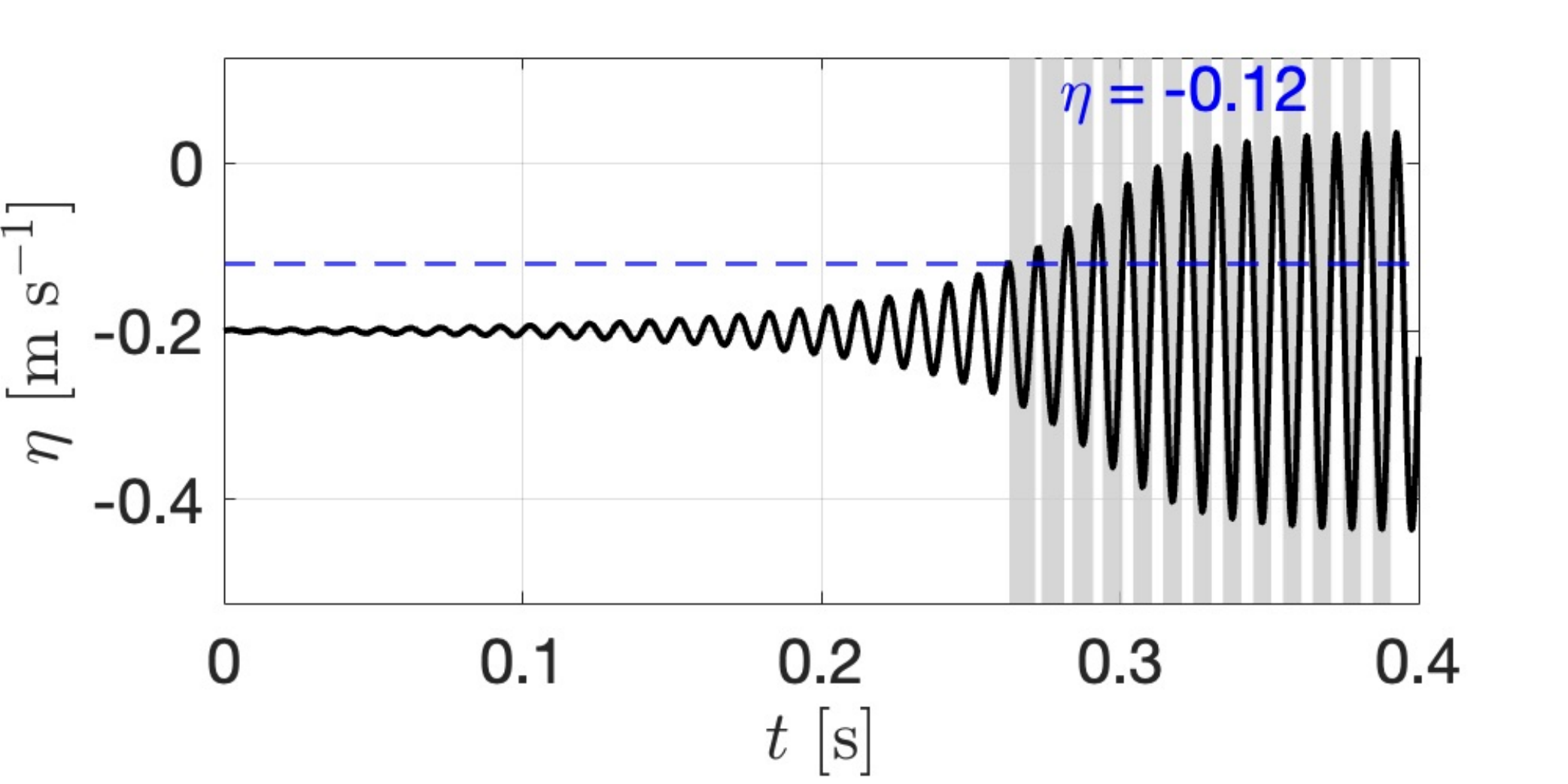}}
\subfloat{\includegraphics[width=0.3\linewidth]{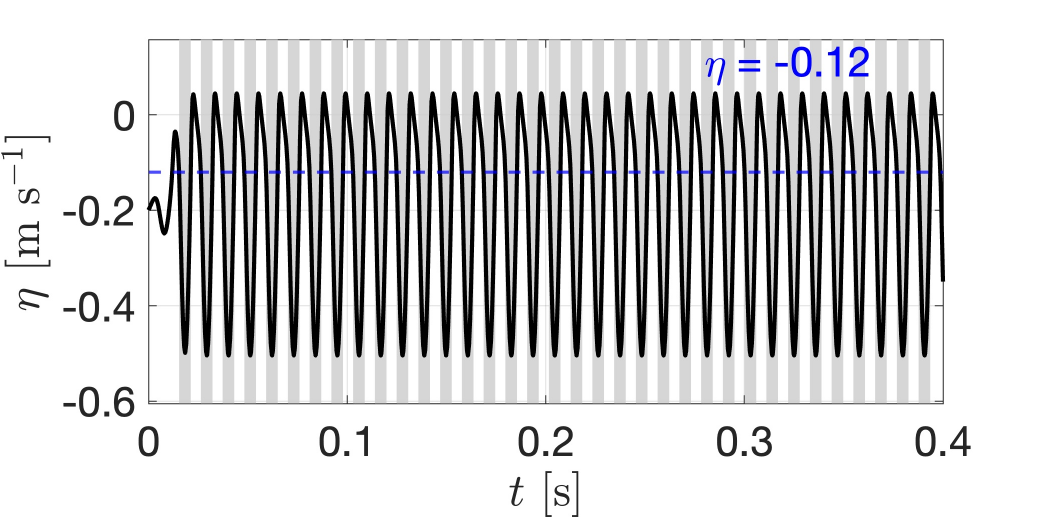}}
\subfloat{\includegraphics[width=0.3\linewidth]{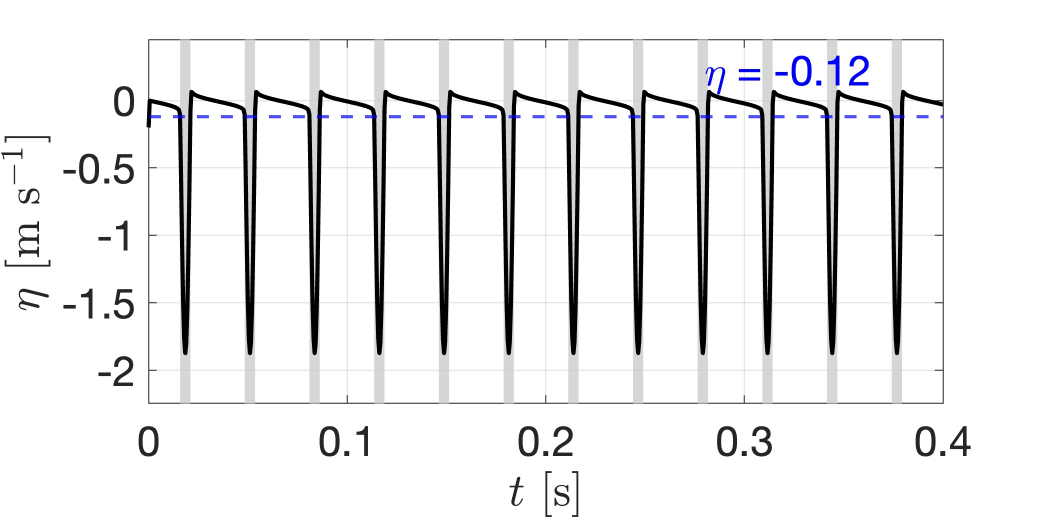}}
    \vspace{-4mm}
    \caption{First and second rows: Simulation results. Third and fourth rows: ODEs loss distribution. Fifth row: Histogram of Hessian eigenvalue density. Sixth row: Loss landscape for PINNs. Seventh row: Loss landscape for PI-DeepoNets. Eighth row: Relative velocity $\eta$ obtained from FDM, with the slip phases highlighted in the shaded regions.}
    \label{fig: result}
\end{figure*}
\vspace{-2mm}
\subsection{Results}
\subsubsection{Zero ICs: PINNs vs. PI-DeepONets}
The first and second rows of Fig.~\ref{fig: result} present the simulation results for zero ICs ($p|_{t=0} = q|_{t=0} = 0$), where we compare PINNs and PI-DeepONets against the Finite Difference Method (FDM) benchmark. The FDM scheme for the bowed mass-spring model follows the implementation in \cite{bilbao2009numerical} and employs an extremely high sampling rate of $4410 $ \si{\kilo \hertz} to ensure accuracy.
Overall, both PINNs and PI-DeepONets perform well for $F_B = 10$ and $F_B = 100$. However, for $F_B = 1000$, the optimization of PI-DeepONets fails, whereas PINNs successfully converge to the desired solution.
When $F_B = 1000$, training the model becomes more challenging, often leading to failure modes in the bowed mass-spring model.  
To mitigate this issue, we employ causal training and a time-marching strategy with a short time duration. Notably, this is the only model that utilizes causal training.

To further assess the accuracy of the results, we present the ODE$_1$ and ODE$_2$ losses in the third and fourth rows of Fig.~\ref{fig: result}. In particular, we compare the losses of PINNs and PI-DeepONets against FDM. Here, FDM refers to the previously mentioned high sampling rate, while FDM-low represents a lower sampling rate (audio rate) of $44.1$ \si{\kilo \hertz}.
PINNs demonstrate better accuracy than PI-DeepONets and can achieve competitive accuracy with FDM at the high sampling rate. Meanwhile, PI-DeepONets can achieve competitive accuracy with FDM at an audio sampling rate. When examining PINNs for $F_B = 1000$, it is worth noting that even FDM at an audio sampling rate can exhibit relatively large local numerical errors, although the results may still be acceptable for sound synthesis purposes.
This suggests that the challenge is not exclusive to PINNs and PI-DeepONets but also affects traditional numerical methods.

The results above intuitively suggest that for $F_B = 1000$, the optimization process is ill-conditioned. To further validate this hypothesis, we analyze the deep learning dynamics using both the Hessian eigenvalue density histogram and loss landscape visualization (for PINNs, we only show the result from the first window network for time-marching). 
The Hessian matrix, denoted as 
$\displaystyle{H =\nabla^2 \mathcal{L}(\mathbf{\theta})_{i,j} = 
\frac{\partial^2 }{\partial_{\theta_i}  \partial_{\theta _j}} \mathcal{L}(\theta)
}, H \in \mathbb{R}^{n_\theta \times n_\theta }$, consists of the second-order partial derivatives of the loss function with respect to the neural network parameters,  where $n_\theta$
is the total number of parameters. It captures the information about the curvature of the loss landscape and can be computed using automatic differentiation.
The Hessian matrix can be decomposed as $H = Q \Lambda Q^{T}$, where
$Q$ contains the eigenvectors and 
$ \Lambda = diag(\lambda_1, . . . , \lambda_{n_\theta })$ is a diagonal matrix of the corresponding eigenvalues.

The Hessian eigenvalue density histograms are shown in the fifth row of Fig.~\ref{fig: result}. 
% (Note that the comments below do not consider the PI-DeepONets, $F_B = 1000$ since the model fails training). 
Pyhessian \cite{yao2020pyhessian} is utilized for the calculation of Hessian matrix.
When analyzing PINNs and PI-DeepONets separately, we observe that as $F_B$
  increases, the maximum eigenvalue also increases. Furthermore, when comparing PINNs to PI-DeepONets, PI-DeepONets generally exhibit higher maximum eigenvalues. Overall, the observed maximum eigenvalues are quite large, indicating a high condition number (ratio of maximum to minimum eigenvalues), which suggests that the optimization problem is ill-conditioned. Moreover, as $F_B$ increases, the level of ill-conditioning also intensifies.

We visualize the loss landscape of PINNs and DeepONets in Fig.~\ref{fig: result}, shown in the sixth and seventh rows, respectively.
The network parameters are perturbed along two directions $\varepsilon_1, \varepsilon_2$: the eigenvectors corresponding to the top two Hessian eigenvalues, resulting in the network parameters as $\theta^* = \theta + \alpha\varepsilon_1 + \beta \varepsilon_2$, with $\alpha \in [-0.5, 0.5], \beta \in [-0.5, 0.5]$.
Moreover, layer-wise normalization is applied, following \cite{li2018visualizing}. 
Sharp minima are consistently observed in all cases. These correspond to regions in the loss landscape where the loss function varies rapidly. Mathematically, this also aligns with the previously shown large Hessian eigenvalues, indicating high curvature.
% Notably, the loss landscape is presented on a logarithmic scale. 
% Additionally, the highest loss values reach magnitudes between $10^4$ and $10^{12}$, which are significantly larger.
% Utilizing a second-order optimizer is suitable for this sharp characteristics of the loss landscape, providing justification for our choice of the SOAP optimizer.
The comparison of the loss landscapes of PINNs and PI-DeepONets reveals that PINNs tend to have sharper minima. This is consistent with the intuition that sharper minima often correspond to poorer generalization performance \cite{keskar2016large}, although this may not hold strictly in all cases. 
 Moreover, this observation is consistent with expectations, given that the goal space $\mathcal{G}$ of PI-DeepONets is significantly larger than that of PINNs.  
In essence, PINNs are not expected to generalize well, as their test data replicates the scenarios used during training. Conversely, the test set for PI-DeepONets contains randomly sampled ICs, which will be detailed later. This indicates that optimizing PI-DeepONets is inherently more complex than optimizing PINNs.

\vspace{-2mm}
\subsubsection{Random ICs: PI-DeepONets}

\begin{figure}[h!]
\vspace{-3mm}
    \centering
    \subfloat{\includegraphics[width=0.3\linewidth]{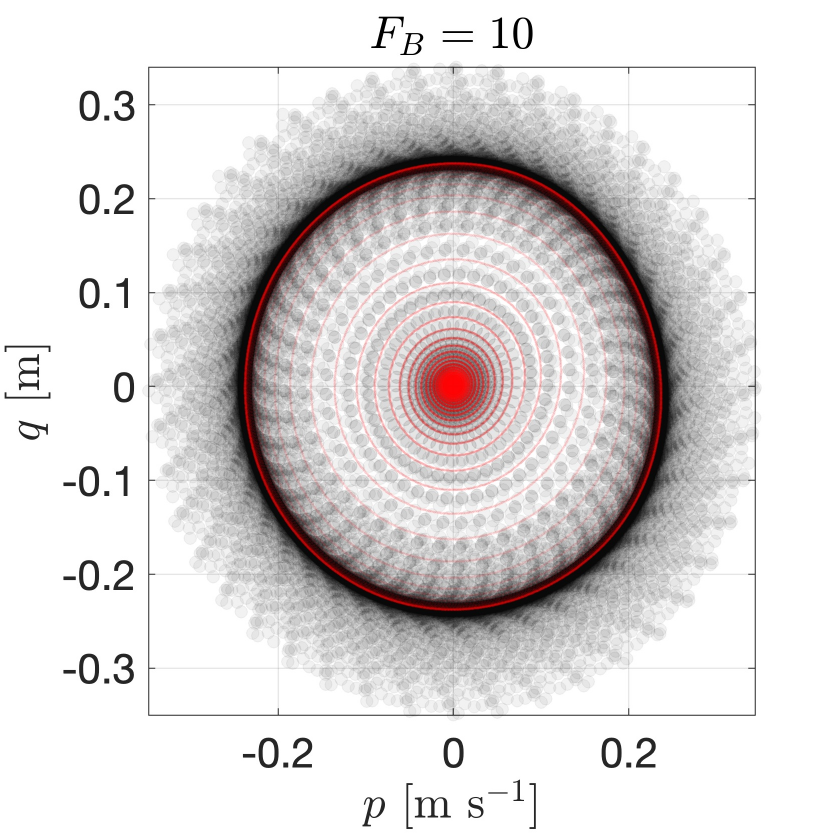}}
         % \hspace{0.1\linewidth}
    \subfloat{\includegraphics[width=0.3\linewidth]{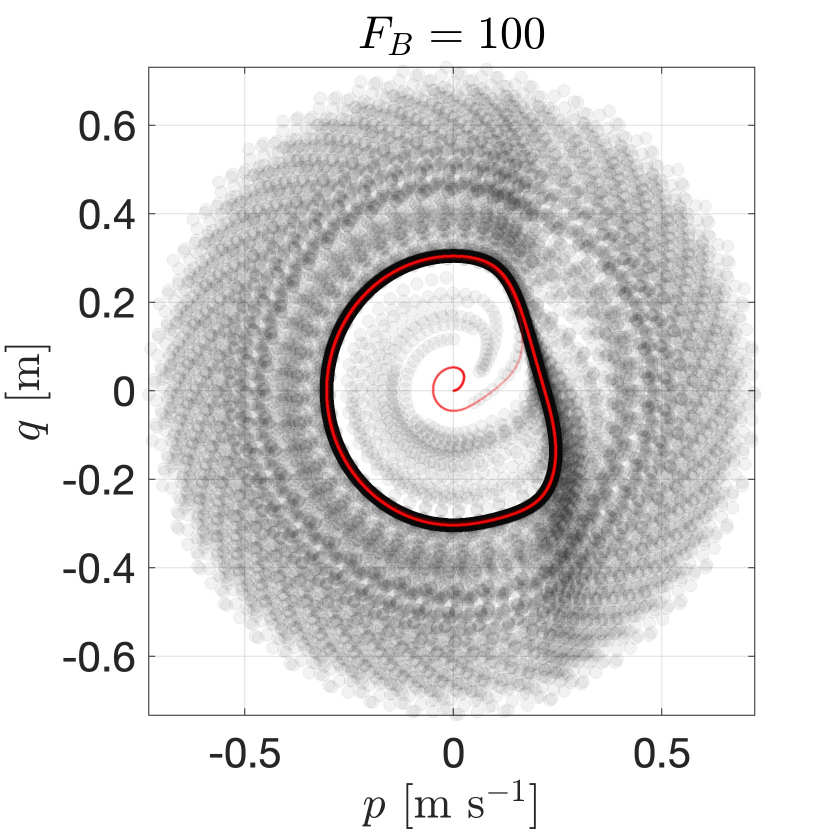}}
    % \hspace{0.1\linewidth}
    \subfloat{\includegraphics[width=0.3\linewidth]{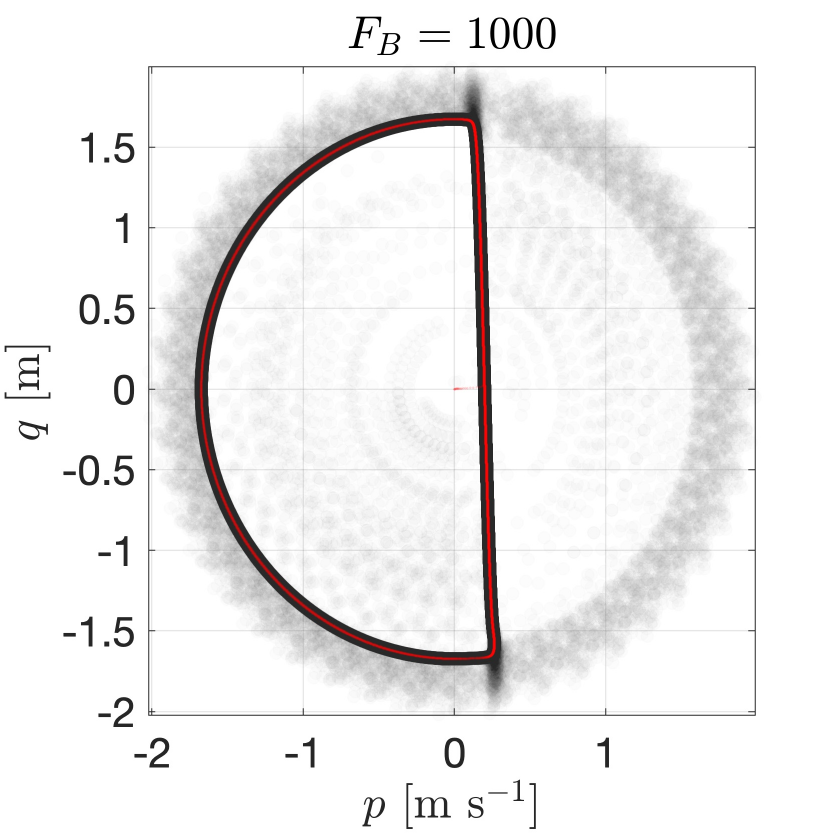}}
    \vspace{-4mm}
    \caption{The solution trajectories of $p$ and
$q$ from the FDM simulation with random ICs. The zero ICs cases are marked in red.}
    \label{fig: demo}
     \vspace{-2mm}
\end{figure}

\sloppy
We evaluate the generalization of PI-DeepONets for different $F_B$ by testing on $100$ cases with uniformly sampled ICs within $ [-s^{p,q}, s^{p,q}] $. Additionally, we ensure that the resulting values of $p$ and $q$ within the time domain stay within the same range. The maximum test time  $t_{max}$ is provided in Table~\ref{tab: acc}. The solution trajectories of the test $p$ and $q$ from the FDM simulation are shown in Fig.~\ref{fig: demo}. 
For each case, the solid black region corresponds to the steady-state solution, while the transient behavior varies with different ICs, resulting in the lighter black spots observed in the figure. We also mark the zero ICs cases in red.
The mean normalized mean square error ($\operatorname{NMSE}$) and normalized cross correlation  ($\operatorname{NCC}$) across $100$ cases are reported in Table~\ref{tab: acc}.
We observe that the predictions for $F_B = 10$ and $F_B = 100$ are almost perfect, while the results for $F_B = 1000$ are notably poor.

\begin{table}[!ht]
\vspace{-1mm}
\captionsetup{skip=0pt} 
 \caption{Prediction results of PI-DeepONets for randomly sampled ICs.}
 \vspace{-3mm}
 \begin{center}
 \footnotesize
 \begin{tabular}{l@{\hspace{2pt}}|l@{\hspace{2pt}}|l@{\hspace{2pt}}|l@{\hspace{2pt}}|l@{\hspace{2pt}}|l@{\hspace{2pt}}}
  % \hline
   $F_B$ & $t_{max}$ [\si{\second}] & $\operatorname{NMSE}(p) $ & $\operatorname{NMSE}(q)  $& $\operatorname{NCC} (p)$ & $\operatorname{NCC}(q)$  \\
  \hline
  10 & 0.4 &$1.70 \times 10^{-8}$ &$1.78 \times 10^{-8}$ & 100.00\% & 100.00\%\\
  100 & 0.2 &  $2.57 \times 10^{-3}$ & $2.19 \times 10^{-3}$& 99.87\% & 99.89\%  \\
  1000 & 0.1 & 2.81 & 1.69 & 5.80\% & 3.37 \%\\
  % \hline
 \end{tabular}
\end{center}
 \vspace{-11mm}  
 \label{tab: acc}
\end{table}

% \setstretch{0.9} 
\subsubsection{Hybrid PI-DeepONets for the failure mode}

The results demonstrate that both PINNs and PI-DeepONets can effectively solve the bowed mass-spring model. However, for $F_B = 1000$, PI-DeepONets fails to converge, while PINNs require time-marching and causal training.
From a physics perspective, the nonlinear bow-string friction interaction is known to induce stick-slip behavior \cite{chaigne2016acoustics}, which defines the bowing characteristics or texture of the sound. 
% \textcolor{red}{todo: For bowed string sound synthesis, perhaps our primary objective is to generate sound that operates within the stick-slip phase. }
In the eighth row of Fig.~\ref{fig: result}, we highlight the slip phases corresponding to various bow force $F_B$ scenarios, using the relative velocity $\eta$ derived from the FDM simulation over the same time duration. The identification of stick-slip phases follows the criteria described in Section~\ref{sec: bow} and Fig.~\ref{fig: phi}, where the slip phase is defined for $\eta \in [-0.12, 0.12]$.
 It is observed that in the bowed mass-spring model, a larger bow force $F_B$ leads to a shorter slip phase, indicating that the system spends more time operating within the highly nonlinear region.
 This, in turn, leads to failure modes. The heightened complexity further challenges PI-DeepONets, making it harder for them to generalize across the broader goal space compared to PINNs.

\begin{figure}[t!]
\vspace{-6mm}
    \centering
    \includegraphics[width=0.7\linewidth]{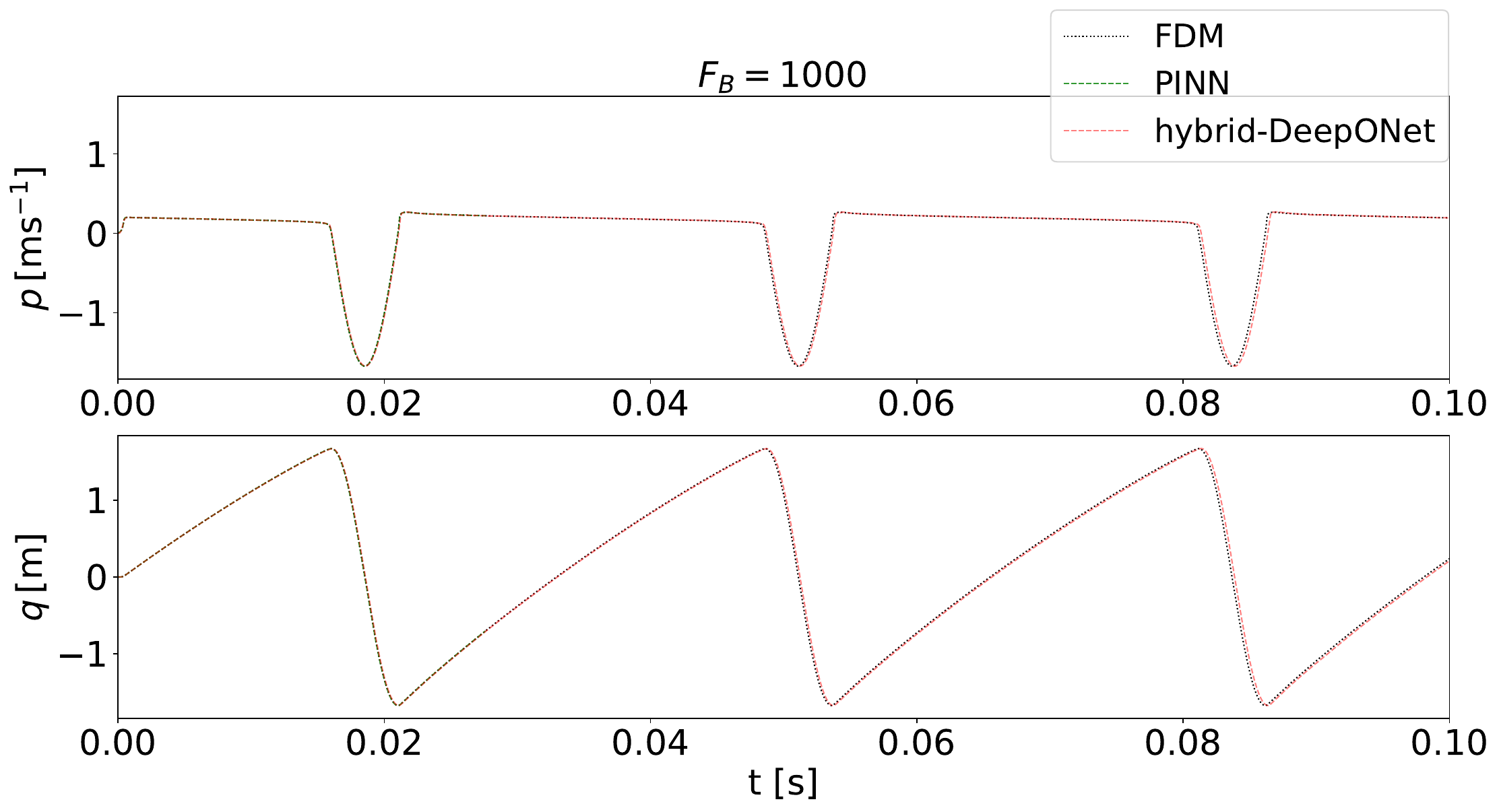}
     \vspace{-4mm}
    \caption{Comparison of simulation results for $F_B = 1000$: FDM vs. PINNs vs. hybrid DeepONets.}
    \label{fig: hybrid}
     \vspace{-7mm}
\end{figure}

 A common approach to mitigating this issue is extending the framework by incorporating data-driven methods. For instance, integrating observation solutions from FDM as an additional term in the loss function term can help guide the network toward the expected solution, transforming the approach into a hybrid unsupervised–supervised training scheme. This idea has been well explored in PINNs research \cite{gopakumar2023loss}, where supervised training has been shown to significantly ease optimization. 
Therefore, we add 

\begin{equation}
{ \mathcal{L}_{{ob}_1} = \frac{1}{N_{ob}}  \big \| \hat{p} - p\big \|,  } \quad
{ \mathcal{L}_{{ob}_2} = \frac{1}{N_{ob}}  \big \| \hat{q} - q \big \|,} 
\end{equation}
where $p$ and $q$ are obtained from FDM with a high sampling rate, given zero initial conditions and $t \in [0,0.1]$. The results are shown in Fig.~\ref{fig: hybrid}. The evaluation metrics comparing the hybrid DeepONet with the high sampling rate FDM under zero initial conditions are: $\operatorname{NMSE}(p) =6.88\times 10^{-3}$,  $\operatorname{NMSE}(q) =1.72\times 10^{-3}$,  $\operatorname{NCC}(p) =99.66\%$ and $\operatorname{NCC}(q) =99.91\%$.
Clearly, the hybrid DeepONets demonstrate accurate predictions. 

%% file: sections/conclusion.tex
\section{Discussion and Conclusion} \label{sec: conc}
\vspace{-1mm}
In this study, PINNs and PI-DeepONets are utilized to solve the nonlinear bowed one-degree-of-freedom mass-spring system. The equation-solving task is formulated as an optimization problem and carried out within the framework of physics-informed deep learning.
Scenarios with different bow forces have been investigated using both PINNs and PI-DeepONets. 
 
 The results show that while PINNs successfully solve the problem across all cases, PI-DeepONets perform well for low bow forces but struggle at higher bow forces. 
 We further analyze the deep learning dynamics by examining the Hessian eigenvalue density and visualizing the loss landscape perturbed along the directions of the dominant Hessian eigenvalues. The presence of large Hessian eigenvalues and sharp minima in the loss landscape suggests ill-conditioned optimization.
This highlights the challenge for purely physics-informed deep learning techniques for modeling the nonlinear bow-string friction interaction.

 Although PINNs can handle scenarios with high bow force, they require additional strategies, such as causal training and the use of short time windows in the time-marching scheme to optimize effectively. While PINNs are limited by fixed ICs and face increasing computational costs as the number of time windows grows, PI-DeepONets excel in generalizing across varying ICs but may fail in highly ill-conditioned scenarios. Our findings suggest that a hybrid approach, combining the strengths of both physics-informed and data-driven methods, can alleviate the limitations of PI-DeepONets for higher bow forces.

% While PINNs are effective at solving the problem, they have limitations, such as being restricted to fixed ICs and experiencing increased computational costs as the number of time windows expands. In contrast, PI-DeepONets demonstrate strong generalization capabilities with varying ICs but may struggle in highly ill-conditioned scenarios. A hybrid approach that combines physics-informed and data-driven methods could help address these challenges.

As this is an initial exploratory study, our primary objective is to investigate the potential of physics-informed deep learning approaches for capturing nonlinearities in string instruments. Therefore, we do not yet demonstrate the use of the trained network for sound synthesis in real-world applications, though we definitely plan to pursue this in future work.
For the deep learning approaches employed in this paper, we train a surrogate neural network to model the physical system, embedding the governing physical laws directly into the training process. While the training phase can be complex, requiring thoughtful strategy design and careful hyperparameter tuning, once trained, the model can be used for inference without explicitly solving the physics. This allows sound synthesis to be performed efficiently by simply running the network, without the need to solve equations in real-time, sample by sample, as is required in traditional numerical methods like FDM. Since the physical dynamics are already captured during training, the inference stage will be significantly faster, making deep learning approaches highly promising for achieving real-time operation, an essential yet challenging goal in physics-informed sound synthesis. We do not include inference in this work, as it falls outside the scope of this paper. However, we are actively working on it and plan to explore it in future studies.

Furthermore, by emphasizing the differences between PINNs and PI-DeepONets in the solution space, we argue that PI-DeepONets are more suitable for sound synthesis applications due to their ability to cover a broader solution space. Considering the nature of musical signals, which typically include transient, steady-state, and decay stages while remaining predominantly periodic, DeepONets are particularly well-equipped to capture these characteristics.
Moreover, DeepONets demonstrate a strong ability to train a single model capable of solving multiple systems governed by the same equation with varying parameters, making them highly adaptable for sound synthesis. 
On one hand, the results underscore the potential of physics-informed deep learning for modeling nonlinear physical processes in musical acoustics. On the other hand, the practicality of a purely physics-informed approach at the current technological level may vary depending on the specific task, as challenges remain in capturing complex nonlinearities. 
To translate these insights into practical sound synthesis applications, a hybrid approach that combines physics-informed and data-driven methods presents a more viable and effective path forward.
Looking ahead, the next step will be to train a hybrid model that incorporates a variety of physical parameters as input, further enhancing its flexibility and applicability.